# Rebuilding Startups: An Empirical Study on Remote Work and Skill Complementarity

**Zixi Lei**
University of Texas at Austin
IROM Department, 2110 Speedway
Austin, TX 78705
zixilei_liz0119@utexas.edu

**Xiaomeng Chen**
University of Pittsburgh
Katz School of Business, 3950 Roberto Clemente Dr
Pittsburgh, PA 15260
xchen@katz.pitt.edu

**Wen Wen**
University of Texas at Austin
IROM Department, 2110 Speedway
Austin, TX 78705
wen.wen@mccombs.utexas.edu

**Andrew Whinston**
University of Texas at Austin
IROM Department, 2110 Speedway
Austin, TX 78705
abw@uts.cc.utexas.edu

## Abstract

Remote work is a common practice of workplace flexibility enabled by digital infrastructure, yet its implications for firms' talent composition and organizational structure remain underexplored. This question is especially important for startups, which often operate with constraints in accessing talent through local labor markets. We focus on skill complementarity, a dimension of talent composition that is critical to startups' early-stage development. By integrating LinkedIn job postings, worker skills, and worker mobility data, we study the impact of remote work on skill complementarity in startups. We find that higher remote work intensity increases firm-level skill complementarity, primarily by helping startups recruit new workers whose skills better match the firms' hiring needs and diversify from the skill sets of existing employees. We further show that this shift in workforce composition carries organizational consequences: remote work is associated with greater organizational hierarchy through increased skill complementarity. These findings suggest that, for startups, remote work is not merely a flexibility policy but a hiring mechanism that helps young firms assemble complementary talent despite resource constraints. At the same time, this may push startups toward more hierarchical structures, which serve as formalized coordination mechanisms to effectively achieve the value of a more complementary workforce.

## 1. Introduction

Remote work is a common practice of workplace flexibility enabled by the modern digital infrastructure, allowing firms to coordinate work and source talent independent of physical location. Digital resilience literature views remote work as an IT-enabled adaptive capacity that helps firms and workers withstand external disruptions (Boh et al. 2023; Bai et al. 2021; Adams-Prassl et al. 2022; Hou et al. 2024). However, there has been considerable debate on whether remote work should be a long-term management decision, given mixed evidence on its organizational effects and hiring decisions (e.g. Gibbs et al. 2023; Choudhury et al. 2026; Yang et al. 2022). Most prior research has examined how remote work affects performance-based outcomes, such as productivity and innovation, and has focused predominantly on established organizations (e.g., Bloom et al. 2015; Choudhury et al. 2021). In contrast, much less is known about how remote work can be used as a hiring choice to influence talent composition and organizational structure within firms, particularly in startups.

Understanding how remote work shapes talent composition in startups is important for several reasons. Compared with established firms, startups typically operate with limited resources, small teams, and fragile organizational structures, where work routines are still being formed (Beckman et al. 2007). Work in startups also tends to involve greater interdependent tasks and close collaboration across team members (Stinchcombe 1965; Eisenhardt and Schoonhoven 1990; Baker and Nelson 2005). These features make startups especially sensitive to changes in the types of talent they can assemble. In this study, we focus on one specific dimension of talent composition within startups, i.e., skill complementarity, which refers to the extent to which workers' skills interact synergistically so that their joint contribution exceeds the sum of their individual contributions during collaborative work (Milgrom and Roberts 1995; Basu et al. 2024; Ennen and Richter 2010).

Motivated by these observations, we propose our first research question: *how does remote work influence skill complementarity among workers in startup firms?* Answering this question can offer useful insights into how startups can assemble teams with complementary skill sets, which are a critical source of innovative capability. Teams with complementary skills bring together distinct knowledge bases and problem-solving perspectives, enabling young organizations to navigate uncertainty, identify novel opportunities, and generate the kind of recombinant innovation that drives growth (Eisenhardt and Schoonhoven 1990; Hellmann and Puri 2002; Lee et al. 2004). However, assembling such teams is often more challenging for startups than for established firms because startups face systematic barriers in accessing and attracting workers with diverse and complementary skills.

Moreover, coordination costs rise when managing a more complementary and diversified team, driving organizations to adopt effective integration mechanisms to overcome the friction and achieve the value of a complementary workforce (Basu et al. 2024). Organizational hierarchy — by centralizing decision-making, clarifying authority, and providing channels for resolving conflict across interdependent roles — is one of the principal mechanisms through which startups can address coordination cost associated with increased complementarity (Lawrence and Poliquin 2023; Puranam et al. 2012). Accordingly, our second research question can be posited as follows: *How does remote work influence organizational hierarchy through its impact on skill complementarity in startup firms?*

Building on the prior literature, our key argument is that by relaxing geographic and financial constraints, remote work enables startups to recruit workers with a more diversified range of skills and better match those skills to their highly interdependent tasks, thereby enhancing skill complementarity. We further argue that remote work leads to greater startups' organizational hierarchy by increasing skill complementarity within startups. Specifically, as remote work enables startups to assemble workers with more complementary skills, the resulting increase in task interdependence and coordination complexity may motivate firms to adopt more hierarchical structures to facilitate decision-making.

To test these predictions, we construct a firm-level panel dataset by integrating LinkedIn job postings, user skill sets, and worker mobility data. We restrict our analysis to startup firms with 1–50 workers and those founded after 2021. We focus on the post-pandemic period because we are interested in understanding how remote work can be used as a strategic hiring choice rather than adjustments driven by pandemic-related shutdowns.

To measure the level of skill complementarity among workers at the firm level, we adapt a network-based approach based on Neffke (2019). We identify skill pairs that are distinct to each other but frequently used together to accomplish work among startups in the same industry. These skills pairs suggest high synergy and value co-creation as they co-occur frequently industry-wide, yet are not substitutable, thereby indicating strong complementarity. Our approach extends prior literature on measuring skill complementarity (Neffke 2019) in two important ways. First, we construct the skill network at the startup level, capturing complementarity demand specific to startups within the same industry. Second, we define complementarity at the skill level, allowing us to fully exploit granular worker-skill data, with workers permitted to hold multiple skills for a more accurate representation of skill co-occurrence.

To provide empirical evidence on how remote work influences skill complementarity and the subsequent organizational structure, we employ a difference-in-differences design that incorporates Rolling Entry Matching (REM) to construct comparable treatment and control groups. Our analysis sample comprises of 2,286 startups after REM, with 29,586 observations from the first quarter of 2021 (2021Q1) through the second quarter of 2025 (2025Q2).

Our empirical results suggest that remote work increases skill complementarity among workers in startups. Specifically, a one-standard-deviation increase in the intensity of remote job postings leads to 13.4% increase in a startup's workforce skill complementarity score, relative to the sample average. This finding is robust across a range of robustness checks, including lagged specifications, instrumental variables estimation, pre-trend analysis, and an alternative measure of skill complementarity. To further uncover the underlying mechanism, we show that remote work increases skill complementarity among workers in startups by enabling them to successfully recruit new workers whose skill sets closely match their hiring needs and diversify the existing skill base of incumbent workers.

Furthermore, we find a positive effect of remote work on organizational hierarchy through skill complementarity. Follow Lee (2021), we employ a measure of organizational flatness, where lower values indicate taller and more hierarchical structures. We find that the intensity of remote job postings is negatively associated with organizational flatness, indicating that greater remote work leads to more hierarchical structures. Moreover, this relationship is partially mediated through an increase in skill complementarity.

Our research makes several contributions to the literature. First, our study is among the first to examine how remote work influences skill complementarity among workers in startup firms. In doing so, we contribute to the IS and economics literature on IT-enabled remote work, which has largely examined how digital technologies shape outcomes for workers already inside the firm (e.g., Bloom et al. 2015; Torres and Crossler 2025; Hou et al. 2024; Choudhury et al. 2021). Prior work also treats telework as a resilience mechanism during disruptions, showing that firms and workers with greater capacity to work remotely are more resilient to economic and societal shock (Bai et al. 2021; Adams-Prassl et al. 2022; Papanikolaou and Schmidt 2022), and that the benefits of this capacity depend on complementary digital infrastructure (Dettling 2017; Hou et al. 2024). We extend this literature by shifting attention from how remote work affect the productivity of an existing workforce to how remote work, as a firm-level hiring choice, changes which workers a startup can access in the first place, and how this reshapes the firm's skill composition. These changes are consequential because they affect how young firms acquire, combine, and coordinate human capital resources, with important implications for their long-term growth and development. By highlighting these organizational consequences, our findings offer a new perspective on the debate over whether remote work should be adopted as a sustained management strategy.

Second, we contribute to the literature on skill complementarity by shifting attention from its consequences to its organizational antecedents. Prior research has established that skill complementarity is a key source of value creation, enabling firms to combine individual capabilities into collective human capital resources that generate performance advantages (Ployhart and Moliterno 2011; Ployhart et al., 2014). Related work shows that skill complementarity shapes employees' wages, tenure, labor mobility, and career outcomes (Anderson et al., 2017; Neffke 2019; Stephany and Teutloff 2024). Yet relatively little is known about how firms develop such complementarity in the first place. We address this gap by identifying remote work as an important organizational antecedent through which startups can access workers with more diverse skills and better match those skills to interdependent tasks, thereby fostering greater skill complementarity within the workforce.

Our findings also offer important implications for startups. Because early-stage firms operate under resource constraints and each hiring decision is crucial and could have lasting consequences (Stinchcombe 1965; Baker and Nelson 2005), they could leverage remote work as a strategic hiring tool to access workers whose skills fit with their interdependent tasks and growth needs. However, startups should recognize that the benefits of more complementary teams often come with greater coordination demands. As startups adopt remote work and recruit workers with more complementary skills, they may need to introduce more formalized coordination mechanisms such as additional hierarchical structures to effectively integrate increasingly interdependent work. Consequently, these structural adjustments toward greater hierarchy may not necessarily be viewed as inefficiencies. Rather, they may represent a productive organizational response that enables startups to capture the value of a more complementary talent base, improve coordination, and build a stronger foundation for growth and execution.

## 2. Literature Review

### 2.1. Remote Work

Remote work refers to arrangements in which workers perform their jobs away from a centralized workplace, enabled by digital communication and collaboration technologies. (Olson 1983, Ellison 1999). It spans a spectrum from fully remote (no mandated on-site presence) to hybrid arrangements (regularly scheduled on-site interaction). As remote and hybrid work arrangements become increasingly prevalent, this topic has attracted significant scholarly attention.

A body of existing literature in IS and economics examines how IT-enabled remote arrangements affect worker productivity and performance. Early evidence suggests that remote work can improve productivity and job satisfaction under certain conditions. For example, Bloom et al. (2015) study a randomized work-from-home experiment in a setting where tasks are relatively standardized with limited need for coordination with coworkers. They find significant improvements in worker productivity and retention. Moreover, prior work shows that technology-mediated interaction substitutes for face-to-face collaboration (Dennis et al. 2008) and "virtuality" shapes trust and knowledge sharing within teams (e.g., Staples and Webster 2008). Prior work also treats telework as a resilience mechanism during disruptions, showing that firms and workers with greater capacity to work remotely weathered the COVID-19 shock better (Bai et al. 2021; Adams-Prassl et al. 2022; Papanikolaou and Schmidt 2022), and that the benefits of this capacity depend on complementary digital infrastructure (Dettling 2017; Hou et al. 2024).

Other studies suggest that the impact of remote work depends on the nature of the task. Dutcher (2012) finds that remote work improves performance on creative tasks but reduces performance on routine tasks. More recent work further emphasizes how the modality of remote work shapes outcomes. Studies on work-from-anywhere policies show that geographic flexibility can increase productivity among knowledge workers by allowing workers to select environments better suited to focused work (Choudhury et al. 2021), while research on hybrid arrangements suggests that reduced physical visibility may influence mentoring opportunities, collaboration patterns, and promotion outcomes (Choudhury et al. 2026).

While these studies provide important insights into the performance implications of remote work such as productivity, job satisfaction, and career advancement, they pay less attention to how remote work reshape firms' talent composition and organization structures. Our study contributes to this literature by specifically examining how remote hiring improves skill complementarity of startups' workforce.

Second, most prior research on remote work examines its effects in established firms with relatively stable organizational settings. In contrast, research examines how remote work operates in startups is relatively scarce. This gap is important because startups differ from established firms in several fundamental ways. Startups typically operate with limited resources, small teams, and fragile organizational structures, where coordination mechanisms and work routines are still being formed (Baker and Nelson 2005; Beckman et al. 2007). Work in startups also tends to involve greater uncertainty, with tasks that are often interdependent and require close collaboration across team members. As a result, remote work may have distinct implications for startups. In our study, we contribute to the remote work literature by shedding light on how remote work influences team composition and organization changes in early-stage startup firms.

### 2.2. Skill Complementarity

Skill complementarity captures a foundational principle in economics and organizational theory: multiple elements of a system jointly produce greater value than the sum of their isolated contributions. At its core, complementarity reflects a super-additive production relationship in which the presence of one element increases the marginal returns of another (Milgrom and Roberts 1995).

In economics, there are two types of frameworks used to detect complementarities (Brynjolfsson and Milgrom 2013). One framework identifies complementarities through a production function that models how different inputs interact to produce systematic performance differences (Krusell et al., 2000; Lindquist 2004). Recent work extends the logic from broad input categories, such as capital and skilled labor, to more granular skill portfolios. For example, Nath et al. (2026) use the framework to examine skill complementarities among AI, IT development, and IT operation skills. The other framework identifies the correlations or co-occurrence in practices (Arora and Gambardella 1990). Significant co-occurrence in practices suggest synergistic interactions among practices. Our study follows this framework of complementarity but shifts the focus to skill complementarity among workers inside firms, where complementarity captures synergistic interactions among individuals whose skill contributions reinforce one another during collaborative work (Basu et al. 2024; Ennen and Richter 2010).

Our perspective also aligns closely with the strategic human capital (SHC) literature, which emphasizes human capital is not merely an attribute of individual workers but a collective asset formed through the ways workers' knowledge, skills, and abilities (KSAs) interact within organizations (Ployhart and Moliterno 2011; Ployhart et al., 2014). As a result, the interdependencies created through the division of labor form the basis for understanding human capital as fundamentally interactive, making complementarity a central mechanism through which human capital contributes to performance advantages, firm specificity, and value capture (Campbell et al. 2012; Wright et al., 2014; Ployhart et al., 2014).

Building on this theoretical foundation, recent empirical studies have begun to model skill complementarity via skill co-occurrence directly. They use network-based methods to conceptualize skills as embedded in systems of related capabilities, where the value of any skill depends on how it links to other skills within the broader skill architecture of the firm or labor market (Nedelkoska et al., 2019; Agrawal et al., 2015; Neffke 2019; Alabdulkareem et al. 2018). Skills that sit at the intersection of multiple relationships carry greater productive potential (Agrawal et al., 2015). In particular, Neffke (2019) develops measures of co-worker complementarity and shows that workers exposed to more complementary colleagues experience higher wages, longer tenures, and faster career advancement. This aligns with broader evidence that a worker's outcomes depend not only on individual human capital but also on how well that capital fits with surrounding team members. Evidence from English NHS hospitals also shows that doctors' retention rises when complementary nurses are retained (Moscelli et al., 2025).

However, most empirical research focuses on measuring complementarities or examining the employment outcomes for complementary workers, such as workers' productivity, career outcomes, and mobility. Little is known about organization decisions that help build a complementary workforce. SHC research has identified several mechanisms through which firms develop human capital, including staffing and selection practices, training investments, and performance management (Wright et al., 2014). However, these studies focus on investments in human capital broadly rather than on the organizational arrangements that help firms assemble complementary workforce. This gap is particularly salient in startups, where each hiring decision is crucial in shaping the company's development trajectory (Choi et al. 2025, Rocha and Grilli 2024, Castellaneta et al. 2026). Our study contributes to these streams of literature by showing the effectiveness of remote work on helping startups to recruit a complementary workforce and the corresponding organizational adjustments due to such workforce changes.

## 3. Hypothesis Development

In organizational settings, *skill complementarity* captures the synergistic interactions among individuals whose skill contributions reinforce one another during collaborative work (Basu et al. 2024; Ennen and Richter 2010). Accordingly, complementarity becomes a critical mechanism through which firms generate performance advantages. However, startups often struggle to assemble teams with meaningful skill complementarity. In this section, we first discuss the constraints startups face and how remote work can broaden the accessible range of diverse talents by relaxing these constraints. Then we discuss how remote work affects firm-level complementarity through enabling startups to recruit more diversified workers. Last, we discuss the adjustment in organizational structures following the improved complementarity.

**3.1. Remote Work and Skill Complementarity**

Remote work may relax geographical constraints, enabling the startups to reach a wider range of workers. A substantial literature documents that knowledge is geographically localized, with ideas and technical expertise clustering within innovative regions (Jaffe et al. 1993; Almeida and Kogut 1999; Song et al. 2003). In these innovative clusters, local labor market is composed of workers trained in closely related skills and specialized technological competencies, leading to a relatively homogenous regional talent pool. While this concentration fosters knowledge spillovers for local firms, it simultaneously restricts access for startups *outside* these clusters, confining them to thin regional pools.

At the same time, startups *within* these clusters confront a different constraint: colocation intensifies competition for talent. Many startups strategically co-locate with large firms to benefit from agglomeration economies, such as knowledge spillovers, access to networks, and ecosystem advantages (Shaver and Flyer 2000; Guzman 2024; Sorenson et al. 2021). Yet large firms typically offer higher wages, stronger employer branding, and more stable career prospects, all of which increase the difficulty for startups to secure high-skilled talent (Bessen et al. 2023). Hence, whether inside clusters where colocation intensifies competition, or outside them where localization restricts access, startups face limited reach to diverse human capital. By removing these geographic restrictions, remote work "thickens" the talent market available to each firm (Aksoy et al. 2022; Choudhury et al. 2021; Hsu and Tambe 2025), particularly for those located in regions with thin local supply.

Startups also face financial constraints to compete for talent. Startups often need to pay wage premiums to attract high-ability workers (Burton et al. 2018; Kim 2018; Sorenson et al. 2021) and frequently struggle to hire high-ability workers (Roach and Sauermann 2024). Remote work helps relax this constraint by offering non-pecuniary benefits: greater temporal flexibility (Angelici and Profeta 2024), reduced commuting burdens (Aksoy et al. 2022), lower living costs and tax burdens (Akan et al. 2025), and improved geographic satisfaction (Choudhury et al. 2021). Workers often value these benefits highly and are even willing to accept wage cuts in exchange for full-time remote options (Mas and Pallais 2017). By offering remote work, startups can thus present a more attractive overall employment package even when they cannot match salaries with large firms, widening the range of candidates they can feasibly recruit.

While remote work can help startups reach a more diverse pool of candidates, such access does not automatically translate into assembling a workforce with greater skill complementarity. Diversification becomes complementary only when workers' distinct skills fit together in ways that improve the firm's capability base. We argue that startups are especially likely to convert diversified hiring into complementarity for two reasons.

First, startups are characterized by resource constraints, high task interdependence, and elevated risk, meaning that each hiring decision can substantially alter the firm's trajectory (Rocha and Grilli 2024; Sorenson et al. 2021). Unlike large or established firms, which often possess greater organizational slack and can maintain overlapping resources as buffers, startups typically cannot afford redundant talent or duplicated capabilities across workers (Bradley et al. 2011). Therefore, they are more incentivized to select diversified skills that contribute synergistic value, such as capturing emerging opportunities or enabling knowledge recombination (Cohen and Levinthal 1990), to offset increased management costs usually associated with a more diversified team. Thus, diversified hiring is often translated into a deliberate effort to improve team-level synergy within startups.

Second, startups operate with high task interdependence (Beckman et al. 2007; Brattström 2024). Unlike larger firms, where roles are more likely to be modularized and separated across departments, startup roles are often tightly connected. Under such conditions, diversified skills are valuable only when they can be integrated into the firm's interdependent workflows. Therefore, the type of diversification startups seek should be diversification that maps onto interdependent tasks and creates mutually reinforcing contributions. When remote work enables startups access to more diversified candidates, the workforce skill complementarity increases if these new hires' distinct skills match with startups' interdependent workflows.

In summary, by relaxing both the geographic and financial constraints that startups face in hiring, remote work enables them to reach a more diversified pool of talent. As a result, startups can select suitable talents to match their demand for complementary skills, enhancing the overall complementarity of their workforce. Thus, we hypothesize:

*Hypothesis 1 (H1): Remote work increases skill complementarity within startups.*

### 3.2. Remote Work and Hierarchy Formation Through Enhanced Skill Complementarity

We further examine how startups reorganize themselves in response to changes in skill complementarity as a result of remote work. As startups begin to hire workers with more differentiated skills, the coordination demands required to integrate these capabilities rise accordingly (Becker and Murphy 1992; Garicano 2000; Lee 2021). We study organizational hierarchy as a key outcome of this process because it captures how startups adjust their managerial structure to accommodate coordination needs arising from improved complementarity. In this sense, organizational hierarchy provides a direct indicator of whether startups respond to greater skill complementarity by introducing more managerial layers. It is therefore an important outcome for understanding how remote work may reshape not only who startups hire, but also how they organize work internally to realize the value of those hires.

While remote work increases the likelihood that firms assemble teams with complementary skills, coordinating such capabilities can be particularly challenging for startups. As nascent firms, they have not yet developed formalized routines, standard operating procedures, or process codification (Stinchcombe 1965; Sine et al. 2006; Cohen et al. 2019). They also often lack informal coordination devices, such as shared norms, strong culture, and dense intra-firm networks (McEvily et al. 2014; Meier et al. 2019). Consequently, workers with complementary skills face a higher risk of conflict and inefficiency unless an integrating mechanism exists to coordinate their efforts.

This absence of coordination infrastructure creates a direct need for a mechanism that can align contributions, resolve disagreements, and integrate interdependent tasks. To meet this need, startups may introduce hierarchical layers that centralize decision-making and establish clearer lines of authority. Organization theory has long argued that as firms divide labor into differentiated roles, they must also develop mechanisms to integrate those roles toward common goals. Otherwise, uncertainty and task interdependence make coordination increasingly difficult (e.g., Puranam et al., 2012). Hierarchy is one of the principal mechanisms which firms can use to address this problem. Research has shown that firms are more likely to expand hierarchy when knowledge scope increases, because broader and more differentiated knowledge bases raise coordination costs and require additional structure to manage them (Lawrence and Poliquin 2023). By providing direction and conflict-resolution channels, a more hierarchical structure may enable startups to manage task complexity and maintain coherent progress across complementary workers, allowing the value generated by complementary skills to be effectively realized. Thus, we hypothesize:

*Hypothesis 2 (H2): Remote work leads to greater organizational hierarchy by increasing skill complementarity within startups.*

## 4. Empirical Approach

### 4.1. Baseline Model Specification

To investigate the effect of remote work on worker complementarity, we employ a difference-in-differences (DID) framework to conduct the analysis. Rather than relying on a binary indicator, our setting features a continuous and time-varying measure of remote intensity that changes within firms over time. This approach is well suited to our setting, as the treatment does not occur as a discrete switch but instead operates with varying degrees of intensity across firms. Continuous treatment variation allows us to capture heterogeneous exposure to remote work and to examine "dose-response" [1] relationships between remote intensity and workforce skill complementarity. Furthermore, treatment timing is staggered across firms: some firms adopt remote work earlier, while others adopt later or never adopt it.

We classify firms that introduce remote work at any point during the 18-quarter study period as treated firms, and firms that never adopt remote work as the control firms. To ensure that treatment and control groups are comparable prior to implementing the DID analysis, we apply Rolling Entry Matching (REM) instead of traditional propensity score matching (PSM).

[1] In many DID applications, treatment is not binary but varies in intensity, or "dose." Danaher et al. (2020), for example, use a continuous DID design to study UK piracy website blocking, defining treatment intensity by users' pre-block visits to sites that were later blocked.

Traditional PSM assumes a single treatment start period and matches treated units to control units based on covariates measured at that common baseline. However, this approach is not well suited for settings with staggered treatment timing, where firms adopt remote work in different quarters. REM addresses this issue by aligning the matching process with each firm's treatment entry period (Witman et al. 2019). Specifically, for each control firm, we construct hypothetical treatment entry times and evaluate its pre-treatment characteristics relative to those entry periods. This allows each control firm to serve as a potential comparison for multiple treatment cohorts. We then match each treated firm to the most comparable control firm based on pre-treatment characteristics measured relative to the same entry period.

Following matching, our baseline empirical model analyzes firm-quarter level data, where *i* indexes companies and *t* indexes quarters, as specified below:

$$Skill_Complementarity_{it} = \alpha + \beta_1 \cdot RemoteIntensity_{it} + \gamma_t + \mu_i + \lambda X_{it} + \varepsilon_{it} \quad (1)$$

The key dependent variable $Skill_Complementarity_{it}$ represents the average skill complementarity across employees for firm *i* in quarter *t*. The independent variable $RemoteIntensity_{it}$ captures the cumulative intensity of remote work, measured as the percentage of remote job postings relative to all postings for firm *i* as of quarter *t*. The coefficient of primary interest, $\beta_1$, captures the effect of remote intensity on skill complementarity. We include firm fixed effects $\mu_i$ and quarter fixed effects $\gamma_t$. The vector $X_{it}$ captures time-varying firm-level control variables, including the number of job postings and number of workers in each quarter.

**4.2. How Remote Work Affects Organizational Hierarchy Through Skill Complementarity**

To further explore whether remote work influences organizational outcomes through its effect on workforce skill complementarity, we employ a formal mediation analysis following the standard four-step approach (Baron and Kenny 1986). Specifically, we investigate whether skill complementarity serves as a mediator linking remote work adoption to organizational flatness.

In the first step, we estimate the effect of remote intensity on the outcome $Y_{it}$, where $Y_{it}$ denotes organizational flatness for firm *i* in quarter *t.* The total effect is captured by the coefficient *c* in the following specification:

$$Y_{it} = \alpha + c \cdot Remote_Intensity_{it} + \gamma_t + \mu_i + \lambda X_{it} + \varepsilon_{it} \quad (2)$$

Next, we evaluate whether remote work predicts the mediator by following the baseline model specification in Equation (1), which examines the relationship between remote work intensity and firm-level skill complementarity.

We then regress the outcome variable $Y_{it}$ on the mediator alone to determine whether skill complementarity is associated with organizational outcomes independent of remote work:

$$Y_{it} = \alpha + b \cdot Skill_Complementarity_{it} + \gamma_t + \mu_i + \lambda X_{it} + \varepsilon_{it} \quad (3)$$

Assuming the three preceding steps yield significant relationships, we proceed to the final step. Specifically, we include both remote work intensity and skill complementarity in the outcome regression as follows:

$$Y_{it} = \alpha + c' \cdot Remote_Intensity_{it} + b' \cdot Skill_Complementarity_{it} + \gamma_t + \mu_i + \lambda X_{it} + \varepsilon_{it} \quad (4)$$

In this specification, some form of mediation is supported if the effect of skill complementarity captured by coefficient $b'$ remains significant after controlling for remote intensity. Across all four steps, we estimate the models at the firm-quarter level and include firm-level fixed effect $\mu_i$, quarter fixed effects $\gamma_t$, and time-varying controls $X_{it}$.

## 5. Data and Measures

### 5.1. Sample Construction

We construct a panel dataset using data from Revelio Labs, which integrates information from LinkedIn job postings, user skill profiles, and worker position histories. The LinkedIn job-posting data provides detailed job-level information, such as job descriptions, posting dates, and job roles. The position-history data capture worker mobility, job titles, seniority levels, and wages over time, while the user-skill dataset contains self-reported skill sets for individual workers. Combining these sources allows us to characterize both the composition of each firm's workforce and the skills associated with its workers.

We restrict the sample to firms with 1 to 50 workers, a range commonly used in entrepreneurship research to identify early-stage startups (e.g., Puri and Zarutskie 2012). Firm size is cross validated using Crunchbase data.[2] This restriction ensures consistent identification of early-stage firms and allows us to capture organizational dynamics within tightly coordinated and resource-constrained environments. Our panel spans from the first quarter of 2021 (2021Q1) through the second quarter of 2025 (2025Q2), focusing exclusively on startups founded in 2021 or later. Focusing on post-pandemic period allows us to isolate remote work as a strategic organizational choice rather than a temporary response to external shocks (Aksoy et al. 2022; Choudhury et al. 2026).

Building on these available data sources, we construct our panel data to examine how remote work shapes skill complementarity and organizational outcomes at the firm-quarter level. First, we utilize LinkedIn job postings data to measure the adoption and intensity of remote work. By analyzing raw job descriptions, we identify the onset of remote work policies on the firm level through the presence of remote-related keywords. We then define $RemoteIntensity_{it}$ as the share of cumulative remote postings relative to all postings for firm $i$ by quarter $t$. Second, we use the worker skill data to construct our primary outcome, firm-level skill complementarity. The skill data includes multiple levels of taxonomy[3], and we rely on the most granular level to capture fine-grained differences in workers' capabilities. We then translate these worker-level complementarity into a firm-quarter average that captures the firm's overall level of skill complementarity. Lastly, we derive organizational flatness from worker seniority levels, represented through a role taxonomy that distinguishes hierarchical layers within the firm.

---

[2] Because Revelio relies primarily on LinkedIn profiles, its employee counts may understate a firm's actual workforce. Thus, we cross-validate firm size using Crunchbase by comparing employee counts from Revelio with the corresponding employee ranges reported in Crunchbase (e.g., 1–10, 11–50 employees). We therefore restrict our sample to firms whose Revelio-based employee counts are consistent with the employee ranges reported in Crunchbase.

[3] Revelio Labs employs proprietary algorithms to cluster the universe of skills into discrete taxonomies at varying levels of granularity (e.g., 25, 50, and 75 clusters). In our analysis, we adopt the 75-cluster taxonomy, as it provides the highest level of clustering detail. For instance, a coarse classification at the 25-cluster level may label a skill broadly as "data analysis," whereas the 75-cluster taxonomy distinguishes more specific technical skills such as programming languages (e.g., C++, MATLAB).

After applying rolling entry matching (REM) to address potential confounding factors, the final sample consists of 2,286 startups — 1,143 treatment startups that adopted remote work at some point after they were founded and 1,143 control startups that did not adopt remote work since the founding date until the end of our sample period. Balance checks in Appendix Table A confirm that the treatment and control groups exhibit statistically insignificant differences in observable characteristics prior to the initiation of remote work.

**5.2. Measuring Skill Complementarity**

Our key dependent variable, $Skill_Complementarity_{it}$, is adapted from the measure developed by Neffke (2019). Intuitively, a workforce exhibits high complementarity when its employees possess skills that are frequently used together to accomplish work but provide distinct capabilities. We extend Neffke's methodology by applying it to more fine-grained skill data and constructing industry-year-specific skill networks. Appendix B describes these methodological extensions in detail.

Below, we describe the steps used to construct this measure. We first estimate pairwise relationships between skills based on how frequently they co-occur across co-workers within firms (denoted as synergy) and the extent to which they are interchangeable (denoted as substitutability) at the industry-year level. Using these industry-year-specific skill relationships, we calculate each worker's skill complementarity relative to their co-workers in the focal firm, defined as the portion of the worker's synergy score that is not explained by substitutability. We then aggregate worker-level complementarity scores to the firm level to measure overall workforce's skill complementarity.

**5.2.1. Synergy**

Synergy captures whether certain skill pairs tend to show up together within firms more often than we would expect by chance. We operationalize this idea by constructing a skill network at the industry-year level, where skills are connected if they often co-occur within firms and are possessed by different workers. Formally, let $I$ be the set of all firms. For each firm $i \in I$, let $W_i$, be the set of workers in firm $i$. For each worker $w \in W_i$ , let $K_{i,w}$ be the set of skills of worker $w$. For firm $i$, we want to find a pair of two workers $w_1$, $w_2$ such that $w_1$ has skill $k_1$ and $w_2$ has skill $k_2$, that is

$$w_1, w_2 \in W_i, \quad w_1 \neq w_2, \quad k_1 \in K_{i,w_1}, \quad k_2 \in K_{i,w_2}$$

We can define an indicator $\mathbf{1}$ (.) to check if these conditions hold for firm *i*. Thus, the co-occurrence counts for skill pair ($k_1$, $k_2$) is:

$$N_{k1k2} = \sum_{i \in I} \mathbf{1}\left(\exists w_1, w_2 \in W_i, w_1 \neq w_2 : k_1 \in K_{i,w_1}, k_2 \in K_{i,w_2}\right) \tag{5}$$

Then we use the co-occurrence counts to compute synergy scores:

$$\widetilde{c_{k1k2}} = \frac{N_{k1k2}}{\frac{N_{k1}N_{k2}}{N}} \tag{6}$$

The normalized co-occurrence $\widetilde{c_{k1k2}}$ measures how often the skill pair ($k_1$, $k_2$) appears together within firms, relative to the expected frequency if skills were randomly distributed. Specifically, the numerator $N_{k1k2}$ is the observed count of firms where both skills co-occur. The denominator $\frac{N_{k1}N_{k2}}{N}$ represents the expected co-occurrence assuming independence, where $N_{k1}$ *and* $N_{k2}$ are the number of firms possessing each skill individually, and N is the total number of firms.

This normalization largely controls for differences in skill popularity. Specifically, highly common skills are more likely to co-occur with many other skills simply because they are very popular. For example, skills such as *Microsoft Office* and *Customer service* may be observed together in many firms, resulting in a high raw co-occurrence count $N_{k1k2}$. However, this high count does not necessarily indicate strong synergy. Because both skills are common, $N_{k1}$ *and* $N_{k2}$ are large. As a result, even if the two skills were independently distributed across firms, their expected co-occurrence $\frac{N_{k1}N_{k2}}{N}$ would also be high. Thus, the normalization measure $\widetilde{c_{k1k2}}$ prevents common skills from being classified as highly synergistic merely because they are widely prevalent.

Lastly, the final normalized synergy score $c_{\mathrm{k1k2}}$ is derived from the raw co-occurrence ratio $\widetilde{c_{\mathrm{k1k2}}}$ to produce a bounded measure between 0 and 1. Figure 1a) illustrates the details of construction graphically.

$$c_{k1k2} = \frac{\widetilde{c_{k1k2}}}{1 + \widetilde{c_{k1k2}}} \tag{7}$$

### 5.2.2. Substitutability

While measuring synergy between skill pairs captures how often certain skills co-occur within firms, synergy alone does not fully capture the nature of skill complementarity. This is because some skills may frequently appear together simply because they serve similar or interchangeable roles within a firm, which we refer to as substitutability. To capture this dimension, we construct a second skill network at the industry-year level that measures substitutability between skills based on the similarity of their occupational profiles.

For example, in the healthcare and wellness services industry, skill sets such as {*clinical research*/*clinical trials*/*medicine*} and {*healthcare*/*hospitals*/*healthcare management*}[4] often appear together across workers. However, these skills largely represent similar functional expertise, and workers with these skill sets can often perform similar roles. Such frequent co-occurrence (i.e. synergy) reflects substitutability instead of complementarity and we need to identify substitutability to isolate complementarity from synergy.

Formally, let $E_{rk}$ denote the number of workers in our sample with skill *k* working in role r. For each skill, we construct an occupational employment vector $\mathrm{E}_{rk} = (E_{1k}, E_{2k}, \dots, E_{rk})$, where $r$ indexes different occupations. This vector describes how skill *k* is distributed across occupations.

The substitutability between skills $k_1$ and $k_2$, denoted as $s_{\mathrm{k1k2}}$, is then defined as the Pearson correlation coefficient between their respective skill vectors across occupations:

$$s_{\mathrm{k1k2}} = \mathrm{corr}(E_{rk1}, E_{rk2}) \tag{8}$$

Higher values of $s_{\mathrm{k1k2}}$ indicate that the two skills are concentrated in similar occupations and therefore tend to serve more overlapping roles within firms. Figure 1b) demonstrates the construction of the substitutability score using an illustrative example.

[4] Note that in the Revelio dataset, each "skill" corresponds to an aggregated cluster of related skills rather than a single atomic capability. For example, the skill set {*clinical research/clinical trials/medicine*} captures a bundle of related medical and research-oriented capabilities. More details are provided in section 6.1.1.

#### 5.2.3. Co-worker Synergy and Substitutability

While the skill-level networks capture relationships between pairs of skills at the industry-year level, our primary interest is to assess how workers' skills relate to one another within the same startup. We next evaluate how each worker's skills relate to the skills held by her co-workers in a focal startup at time *t* based on the skill-pair relationship networks for its corresponding industry in that particular year.

For each worker *w* at firm *i* at time *t* (i.e., quarter in our analysis), we compare the skill set of worker *w* with the skill sets of all other workers employed by the same firm in the same quarter. We then compute two key metrics: the co-worker synergy score $C_{wit}$ and the co-worker substitutability score $S_{wit}$.

$$C_{wit} = \frac{1}{|E_{it}| - 1} \sum_{w' \in I_w, w' \neq w} \left( \frac{1}{|k_{iw}| \cdot |k_{iw'}|} \sum_{k_m \in k_{iw}} \sum_{k_n \in k_{iw'}} c_{k_m k_n} \right) \tag{9}$$

$$S_{wit} = \frac{1}{|E_{it}| - 1} \sum_{w' \in I_w, w' \neq w} \left( \frac{1}{|k_{iw}| \cdot |k_{iw'}|} \sum_{k_m \in k_{iw}} \sum_{k_n \in k_{iw'}} s_{k_m k_n} \right) \tag{10}$$

where $|E_{it}|$ is the total number of workers at firm *i* during time *t*, $I_w$ denotes the set of co-workers of worker *w*, $k_{iw}$ and $k_{iw'}$ denote the set of skills held by worker *w* and co-worker *w'*. $s_{k_m k_n}$ and $c_{k_m k_n}$ are the substitutability score and the synergy score between the skill pair ($k_1$, $k_2$), which are calculated previously.

#### 5.2.4. Complementarity

Consistent with prior literature (e.g., Neffke 2019), worker-level skill complementarity is defined as the portion of co-worker synergy that cannot be explained by substitutability. Intuitively, a worker's skills are considered to be complementary to those of her co-workers when they frequently co-occur but remain distinct to each other. For example, in a software firm, a worker with *UX design* skills may be complementary to co-workers with *backend engineering* skills. These skills often appear together but they support different parts of the product development process. In this case, the worker's skills have a complementary relationship with her co-workers' skill.

Following Neffke (2019), we isolate complementarity from synergy score, by modeling the co-worker synergy score $C_{wit}$ as a function of the substitutability score $S_{wit}$ and a residual term $m_{wit}$, expressed as:

$$C_{wit} = \alpha + \beta S_{wit} + m_{wit} \tag{11}$$

The estimated residual term $m_{wit}$ is our focus of interest here, which represents the workers' genuine complementarities among co-workers within the same company at time *t*. A positive $m_{wit}$ indicates that the observed synergy exceeds what would be expected based on substitutability alone, suggesting that the worker possesses genuinely complementary skills relative to their co-workers. Conversely, a negative $m_{wit}$ signals that the observed synergy is lower than expected given the substitutability, implying that skill overlaps generate more substitutability rather than complementarity.

**5.3. Measuring Organizational Hierarchy**

Following the approach in Lee (2021), we focus on organizational flatness as a measurement for hierarchy in our setting. A flatter organization is characterized by fewer managerial tiers relative to total employment, whereas a taller organization exhibits more vertical layers.

More specifically, we construct a firm–quarter–level measure of organizational flatness based on the distribution of workers across hierarchical levels derived from job title classifications. We assign each job title to a hierarchical level using keyword-based rules that capture signals of authority and managerial responsibility. Job titles are mapped into 12 hierarchical levels, ordered from top to bottom: *Owner, president, VP, CEO, C-Suite, Head, Director, Manager, Producer, Lead, Supervisor, and Other.* These levels are derived from qualitative industry sources, validated through practitioner interviews, and operationalized using systematic text analysis of qualitative references (e.g., Kent 2001).

Then, for each firm–quarter, we compute a flatness measure defined as the ratio of the total number of workers to the number of hierarchical levels represented within the firm during that quarter. Formally,

$$Flatness_{it} = \frac{NumberOfEmployees_{it}}{HierarchyLevels_{it}} \tag{12}$$

A lower value of flatness indicates a more hierarchical (i.e., taller) organizational structure, whereas a higher value reflects a flatter structure with broader spans across fewer levels. As a robustness check, we re-estimate our results using alternative job title taxonomies[5] to ensure that our findings are not sensitive to a particular hierarchical classification scheme. As organizational restructuring may not occur contemporaneously with changes in workforce composition, we lag the measurement of organizational flatness by one quarter to capture delayed adjustments in organizational structure, allowing time for internal reorganization processes to materialize.

### 5.4. Descriptive Statistics

Table 1 reports descriptive statistics for the key variables in our analysis at the firm-quarter level (N = 29,586). The average skill complementarity across firms is 0.004 (SD = 0.046), indicating that firms on average exhibit balanced levels of complementary and substitutable skill relationships among coworkers. Remote work intensity has a mean of 0.129 (SD = 0.268), suggesting that approximately 12.9% of job postings are offered remotely.

Firms in the sample are generally small startups, with an average workforce size of 10.5 employees (SD = 9.077), and post approximately 3.6 new job openings per quarter (SD = 82.489). Organizational flatness has a mean value of 9.455 and exhibits meaningful dispersion (SD= 8.256), indicating heterogeneous organizational structures even among relatively young firms. The average annual worker salary is $94,023 (SD = $67,660), with salary values scaled by 1,000 in the table for readability.

## 6. Results

In this section, we first present our empirical estimation and validation results of the skill complementarity measure, followed by analyses examining the impact of remote work on workforce skill complementarity. We then report instrumental variable estimates, mechanism analyses, and supplementary results that further support our main findings.

[5] The Revelio dataset provides multiple job taxonomy classifications at varying levels of granularity. For example, *Role_k150* categorizes positions into 150 discrete role groups, while *Role_k1500* offers a more fine-grained classification with 1,500 distinct categories. In our main analysis, we adopt the *Role_k50* taxonomy to balance interpretability and measurement stability. To ensure robustness, we replicate our analyses using alternative taxonomies (e.g., *Role_k150* and *Role_k1500*) and find that our results remain qualitatively consistent across different levels of aggregation.

## 6.1. Skill Complementarity

### 6.1.1. Descriptive Results on Skill Complementarity

Based on the approaches in Section 5.2, for each industry and year segment, we calculate the synergy score $c_{k1k2}$ and substitutability score $s_{k1k2}$ for every possible skill pair across all startups in our sample economy.[6] Table C1 and Table C2 in Appendix present these measures for three major sectors as illustration examples, listing the highest and lowest synergy skill pairs, as well as the highest and lowest substitutability skill pairs. For instance, in the Healthcare and Wellness Services sector in the year 2023, the highest synergy pairs are {*clinical research/clinical trials/medicine*} and {*medical devices/pharma sales/cardiology*}[7]. This indicates that the skill pair frequently co-occurs within the same firms, suggesting that it exhibits high synergy. By contrast, the lowest-synergy pairs are {*chemistry/molecular biology/biochemistry*} and {*emergency management/physical security/crisis management*}, indicating that this skill pair rarely co-occur together within the same firms, reflecting relatively unrelated workflows.

The substitutability story shows the flip side. Within the IT industries, highest substitutability pairs are {*HTML/JavaScript/MySQL*} and {*Java/Eclipse/Android*}. This indicates that workers possessing these skill pairs tend to be employed in similar roles and thus these two skills are highly interchangeable. By contrast, the lowest-substitutability pairs are {*management/leadership/training*} and {*linux/unix/perl*}. This suggests that these skill pairs map into very different occupational domains and are non-interchangeable.

Figure 2 shows the overall distribution of worker-level skill complementarity score $m_{wit}$ related to co-workers, where positive values indicate that a worker's skills complement those of co-workers and negative values indicate substitutability. The histogram is tightly centered around zero with a modest right tail, indicating that the workforce contains a mix of complementary and substitutive profiles. The slight right tail suggests a smaller subset with pronounced skill complementarity with co-workers, while the left tail indicates some workers whose skills are largely substitutable relative to peers.

[6] Throughout the paper, we use the term "economy" to refer to the startup economy within the same industry, rather than the broader economy that include all firms. It allows us to better capture the complementary skills prevalent in the startups.

[7] In the Revelio dataset, each "skill" corresponds to an aggregated cluster of related skills rather than a single atomic capability. Specifically, each skill label represents a group of closely related competencies. For example, the skill set *clinical research/clinical trials/medicine* captures a bundle of related medical and research-oriented capabilities, while *HTML/JavaScript/MySQL* reflects a cluster of web development skills. As such, our analysis operates at the level of skill bundles, which provides a more realistic representation of how capabilities are organized and deployed in the labor market.

In Figure C1 in Appendix, we plot the average synergy ($C_{wit}$) against the substitutability score ($S_{wit}$) by industry, with the vertical distance from the fitted line indicating average co-worker complementarity. Knowledge-intensive sectors, such as IT Consulting Services and Research and Development, generally lie above the line, consistent with their reliance on heterogeneous expertise to produce more complex products, services, and innovations (e.g., Grant 1996). In contrast, industries like Apparel Retail and Food & Beverage fall below the line: firms in these sectors tend to employ workers who are more interchangeable. Creative and operational sectors such as Culture & Entertainment cluster nearer the baseline, indicating a mixed workforce of complementary specializations and interchangeable roles.

**6.1.2. Wage Prediction**

Prior literature suggests that complementarities enhance productivity by increasing the marginal returns of individual skills (Milgrom and Roberts 1995), which are subsequently reflected in wage premiums. To validate our measure of skill complementarity, we examine its association with firm-level average employee wages and find a positive and statistically significant relationship, as reported in Appendix Table C3. This result is consistent with prior theory and suggests that our measure captures economically meaningful complementarities within firms.

**6.1.3. Alternative Measure of Skill Complementarity**

To further validate our measure of skill complementarity, we construct an alternative firm-level measure $Skill_Complementarity_Alternative_{it}$ based on the network approach developed by Alabdulkareem et al. (2018). The measurement details are reported in the Appendix C.3.

Overall, we find a positive and statistically significant correlation between the main firm-level complementarity measure $Skill_Complementarity_i$ and the alternative firm-level measure

*Skill_Complementarity_Alternative*$_{it}$, [8] providing additional evidence for the validity of our skill complementarity measure. Figure C2 in Appendix compares the distributions of the two complementarity measures. Because *Skill_Complementarity_Alternative* is based on normalized co-occurrence patterns of skills and bounded between 0 and 1, the resulting scores are inherently strictly non-negative, ranging from 0.04 to 0.86 with a mean around 0.45. In contrast, the original measure *Skill_Complementarity* captures the residual difference between observed worker-level synergy and the component explained by substitutability, resulting in a distribution that spans negative and positive values and centered around zero.

**6.2. The Impact of Remote Work on Skill Complementarity**

Table 2 reports the baseline results. We first exclude controls and then include controls for robustness. Across both columns (1) and (2), the coefficient of *RemoteIntensity* suggests that the intensity of remote work plays a critical role in enhancing skill complementarity at the firm level. For example, column (2) shows that a one-standard-deviation increase in the intensity of remote job postings is associated with approximately 13.4% increase in a startup's workforce skill complementarity score, relative to the sample average. This supports our H1 that argues a positive impact of remote intensity on firm-level skill complementarity. The results suggest that remote work facilitates hiring and integrating complementary skill sets to startups, contributing to higher levels of firm-level skill complementarity.

Startups typically make hiring decisions within relatively short time horizons, typically within 30 to 90 days, given their small size and operational urgency. Therefore, in the baseline estimates, we use skill complementarity in the same quarter as remote work intensity, which should leave enough time for changes in remote work intensity to affect firms' skill composition. However, to allow for possible delays between job postings and realized hires, we also estimate lagged models in which skill complementarity is measured in the subsequent quarter. As shown in column (3) and (4) in Table 2, the results remain consistent without and with the inclusion of control variables.

---

[8] This correlation is estimated by regressing *Skill_Complementarity_Alternative*$_{it}$ on *Skill_Complementarity*$_{_it}$, including firm and quarter fixed effects. The coefficient is positive and statistically significant ($\beta = 0.261$, $SE = 0.052$, $p < 0.001$), providing evidence on the validity of our complementarity measure.

### 6.3. Instrumental Variable Estimation

To address the endogeneity issue that both remote work intensity and skill complementarity may be correlated with some omitted unobservable, we implement an instrumental variable estimation to strengthen our causal identification. We construct an instrument based on the remote work intensity of firms that operate in industries most distant from the focal firm in terms of human capital requirements, while being founded in similar years and headquartered within the same metropolitan statistical area (MSA). The construction details of our instrumental variables are reported in Appendix D.

This instrument satisfies the relevance condition because firms adjust organizational practices in response to peer behavior and information spillovers (e.g., DiMaggio and Powell 1983). It also satisfies the exclusion restriction, because peer firms are drawn from the most distant industries in terms of human-capital relatedness (Neffke et al. 2017). The limited overlap in skill requirements and labor mobility makes it relatively unlikely that peer firms directly affect the focal firm's workforce composition. Accordingly, peer remote intensity should affect the focal firm's organizational outcomes primarily through its influence on the firm's own remote-work adoption.

Table 3 reports the instrumental variable (IV) results based on this instrumental variable (denoted as *Peer_Remote_Intensity*). As shown in the first stage in column (1), this instrumental variable is positively and strongly associated with a firm's own remote intensity, suggesting that peer adoption meaningfully predicts firm-level remote practices. The corresponding first-stage F-statistics is 477.912, well above the conventional threshold of 10. These values indicate that the instrument is sufficiently strong in all specifications and alleviate concerns about weak instruments. In the second stage, the predicted remote intensity exhibits a positive and statistically significant effect on firm-level skill complementarity after adding the full set of controls. This finding is consistent with our baseline DID estimates. We further restrict the sample to firms founded within ±5 years of the focal cohort as an additional robustness check and the results remain qualitatively similar.

### 6.4. How Remote Work Affects Organizational Hierarchy Through Skill Complementarity

We next examine whether remote work further affects organizational flatness through skill complementarity, using a mediation framework as discussed in Section 4.2. We begin with the baseline specification estimating the total effect of remote intensity on flatness. Column (1) in Table 4 shows that the coefficient on *Remote_Intensity* is negative and statistically significant, indicating greater reliance on remote work leads firms to adopt more hierarchical structures.

As documented in our baseline results, we have already established that remote intensity positively affects skill complementarity. We now turn to the third stage of the mediation framework by examining whether skill complementarity, serving as the mediator, influences organizational flatness. As shown in Table 4 column (2), the coefficient on *Skill_Complementarity* is negative and statistically significant. This result implies that firms with higher internal skill complementarity among workers exhibit substantially more hierarchical structures. This pattern is consistent with the theorization that as skill complementarities intensify, firms formalize authority and coordination structures to manage interdependent tasks.

Finally, we run the joint specification to test if the effects of skill complementarity on organizational flatness are still significant after controlling remote intensity, as shown in Table 4 column (3). In the joint specification including both *Remote_Intensity* and *Skill_Complementarity*, the coefficient estimate of *Skill_Complementarity* remains negative and statistically significant. To further validate the mediation effect, we conduct a bootstrap mediation analysis (Preacher and Hayes 2008). The bootstrap results confirm that the indirect effect of remote work on organizational flatness through skill complementarity is negative and statistically significant, with the 95% confidence interval excluding zero, providing additional support for the partial mediation effect.[9]

Overall, our results indicate partial mediation: remote work increases skill complementarity, which in turn reduces flatness. At the same time, remote work also retains a total negative effect on organizational flatness beyond the complementarity channel. It suggests that remote work may introduce additional communication and coordination burdens through increasing workforce skill complementarity, which encourages the adoption of more formalized and hierarchical structures.

---

[9] The estimated indirect effect is −0.0086 (bootstrap SE = 0.004, $p = 0.035$). The 95% bootstrap confidence interval is [−0.0167, −0.0006], excluding zero.

To account for potential delays in how workforce composition translates into organizational adjustments, we further include lagged specifications, where organizational outcomes are measured in the subsequent period. The results, as shown in columns (4) to (6) in Table 4, remain consistent.

### 6.5. Mechanism Analysis

As we discussed in section 3.1., remote work enhances skill complementarity by enabling startups access to a more diverse applicant pool and identify new hires whose skills are simultaneously well-matched to the firm's hiring demand such as interdependent tasks. In other words, the underlying mechanism is that remote work increases skill complementarity by recruiting workers whose skill profiles diversify the firm's incumbent workforce and closely match firms' contemporaneous hiring needs. In this section, we provide some empirical evidence that supports this proposed mechanism.

First, we measure *skill fit* by comparing the skills required in firm *i*'s job postings with the skills possessed by newly hired employees within the same quarter *t*. Second, we measure *skill diversification* by comparing the existing skills of incumbent employees with the skills possessed by newly hired employees, similarly within the same firm-quarter. We restrict the analysis to firm-quarter observations with at least one new hire. Both *skill fit* and *skill diversification* measures leverage semantic embeddings generated from a pretrained Sentence Transformer language model. Each skill is first converted into a high-dimensional embedding vector that captures semantic meaning beyond exact keyword matching, and then the similarity between two skills is measured using cosine similarity between their corresponding embedding vectors.

Because both *skill fit* and *skill diversification* may vary systematically over time, we benchmark each measure against the quarterly median across all firms in our sample. We then construct a binary indicator *skill_fit_diversification* which equals one only when both the *skill fit* and *skill diversification* scores are high (i.e., above their respective quarterly medians). This measure essentially indicates whether newly hired employees both closely match hiring requirements and substantially diversify the firm's existing skill base.

We next implement the mediation framework to test the mechanism. First, we confirm the baseline relationship between remote intensity and firm-level skill complementarity. Second, we examine whether remote work intensity leads to higher *skill_fit_diversification*. Third, we test whether new hires' *skill_fit_diversification* binary indicator is positively associated with skill complementarity. Finally, we include both remote work intensity and *skill_fit_diversification* into one specification.

Table 5 reports the estimation results. As shown in column (1), remote work intensity positively and significantly affects the likelihood that startup firms recruit new employees whose skills closely match firm demands and diversify the existing skill base (i.e., the likelihood that *skill_fit_diversification* indicator turns one). Next, as shown in column (2) in Table 5, *skill_fit_diversification* is positively and strongly associated with firm-level skill complementarity. This result suggests that employees who possess high skill fit with the firm's hiring needs and introduce diverse capabilities relative to the incumbent workforce bring complementarity to the startups. Lastly, we include both remote work intensity and *skill_fit_diversification* in the regression in column (3). The persistence of the mediation effect and the attenuation of the remote intensity coefficient are consistent with the partial mediation effect. We further implement the bootstrap analysis with 5,000 replications. The estimated indirect effect of remote work on workforce skill complementarity through high skill fit and skill diversification is positive and statistically significant. The 95% bootstrap confidence interval excludes zero, supporting our argument about this mechanism.[10]

Taken together, our mechanism analysis confirms that remote work improves skill complementary when two conditions are satisfied: new skills map onto the firm's hiring demand (*skill fit*) and genuinely extend the firm's existing capability base rather than duplicate it (*skill diversification*). Our *skill_fit_diversification* indicator operationalizes this joint requirement, and the results confirm that startups benefit from remote hiring not simply by reaching more heterogeneous candidates, but by using that broader reach to identify workers whose skills are well-matched with their hiring needs and also distinct from those of incumbent employees. This is consistent with our theoretical claim that startups, facing limited slack and high task interdependence, are especially incentivized to convert diversified hiring into deliberate complementarity rather than diversity for its own sake.

[10] The estimated indirect effect is 0.00031 (bootstrap SE = 0.001, $p = 0.003$). The 95% bootstrap confidence interval is [0.0001069, 0.0005104], excluding zero.

**6.6. Additional Evidence on the Mechanism**

To provide additional evidence on the mechanism discussed above, we conduct two additional analyses to examine whether the impact of remote work on complementarity is more salient when it provides greater geographic flexibility and when the startups can demand workers with more distinct skills via remote hiring.

First, we distinguish between postings about fully remote work and postings about hybrid work. If geographic flexibility is the channel that drives the effect, we will expect fully remote positions, which offer greater location flexibility, to exhibit a stronger impact on skill complementarity than hybrid remote positions. Consistent with this expectation, Table 6 column (1) shows that the impact of remote work is primarily driven by postings with fully remote work. [11]

Second, we examine whether remote work affects the demand for complementary skills by comparing the skill requirements in new remote hiring postings with the firm's existing skill portfolio. Importantly, this differs from our mechanism analysis in Section 6.5, which focuses on realized hiring outcomes as reflected in the skill profiles of workers who have actually joined the firm.

We operationalize similarity as the cosine similarity between a new job posting's skill vector and the incumbent workforce skill vector in the corresponding firm-quarter using the sentence-transformer language model. Because high similarity indicates potential substitution (adding skills already abundant) whereas low similarity indicates potential complementarity (filling capability gaps), we split the sample at the 50th percentile of the similarity distribution: "low-similarity" postings (≤ 50th percentile) and "high-similarity" postings (> 50th percentile).

[11] We classify job postings into fully remote and hybrid remote categories based on keyword searches within job descriptions. Specifically, we identify hybrid roles using terms that indicate a mix of remote and in-person work (e.g., "hybrid," "partially remote," "remote and in-person," "onsite some days"). All remaining postings that indicate remote work but do not contain hybrid-related terms are classified as fully remote.

As shown in column (2) in Table 6, the effect of remote intensity on skill complementarity is statistically significant and positive among low-similarity postings, whereas the effect is small and statistically insignificant for high-similarity postings. This pattern suggests that only when startups are actively seeking workers whose capabilities are missing in the incumbent team, remote work improves skill complementarity. As a robustness check, we also apply a stricter 75th percentile cutoff to define low- versus high-similarity groups, and the results remain qualitatively similar, as shown in column (3) in Table 6.

Finally, we combine the two dimensions: remote work arrangement (fully remote vs. hybrid) and skill similarity (low vs. high, defined using the 50th percentile cutoff) to conduct a more granular analysis. The results in column (4) of Table 6 indicate that the positive effects of remote work are predominantly driven by postings that are both fully remote and low in similarity. In contrast, hybrid postings and high-similarity postings contribute little to the observed positive effect of remote work on skill complementarity. The results are qualitatively similar if we use the 75th percentile cutoff to define high vs. low skill similarity, as shown in column (5) of Table 6. Overall, this set of results further support our argument that remote work enhances skill complementarity primarily by enabling firms to source distinctive and complementary skills from a much broader talent pool.

**6.7. Robustness Checks**

We conduct three additional analyses to assess the robustness of our findings, with the full specifications and results reported in Appendix E. First, we re-estimate the relationship between remote intensity and skill complementarity using the alternative measure $\textit{Skill_Complementarity_Alternative}_{it}$. Results are reported in Table E1. The estimates remain positive and statistically significant, consistent with those obtained using the baseline measure. This robustness check reinforces our central claim that greater remote intensity enables firms to assemble more complementary workforce skills. The result also suggests that our conclusions are not driven by a particular measure of skill complementarity.

Second, we assess the parallel-trends assumption using event-time specifications with pre-treatment lead terms. Appendix Table E2 shows that the pre-treatment coefficients are statistically insignificant across specifications using one, two, or three leads, suggesting that treated and control firms did not exhibit systematically different trends in workforce skill complementarity before remote-work adoption.

Third, because conventional two-way fixed-effects estimates may be biased under staggered adoption when treatment effects vary across cohorts or over time, we re-estimate the baseline model using the Callaway and Sant'Anna (2021) estimator. This approach compares each adoption cohort with not-yet-treated firms and aggregates the resulting group-time treatment effects. The estimates reported in Appendix Table E3 are quantitatively similar to the baseline results, suggesting that our findings are unlikely to be driven by biases arising from staggered treatment timing or heterogeneous treatment effects.

## 7. Conclusion

### 7.1. Summary of Main Findings

In this study, we examine how remote work shapes the workforce composition and organizational structure of startups. Using a firm-level data that integrates LinkedIn job postings, worker skills, and worker mobility data from 2021 to 2025, we construct a firm-level measure of skill complementarity based on a network-based approach with fine-grained, worker-level skill data. This measure is validated by its positive association with average wages and its convergence with an alternative, occupation-based complementarity measure. We find that greater remote work intensity leads to increases in firm-level skill complementarity, a result that holds under lagged specifications, instrumental variable estimation, and alternative estimation methods. Our mechanism analysis shows that remote work enables startups to successfully recruit workers whose skill profiles both fit the firm's hiring needs and diversify its existing capability base, thereby facilitating the formation of more complementary teams.

Moreover, we show that by assembling a more complementary workforce, remote work further shifts the organization structure of startups. Namely, remote work may lead to greater organizational hierarchy through its effect on skill complementarity. As startups assemble more differentiated and interdependent capabilities via remote hiring, they may also introduce more hierarchical coordination structures to integrate these capabilities and manage the associated coordination demands. Taken together, our results highlight that remote work not only affects who firms hire, but also how they organize and rebuild their internal talent structure.

### 7.2. Theoretical Implications

Our study offers two primary theoretical contributions. First, we contribute to the IS literature on digital resilience by moving beyond the prevailing view of remote work as an IT-enabled response to external disruption or workplace flexibility arrangement (e.g. Boh et al. 2023; Hou et al. 2024; Bai et al. 2021; Adams-Prassl et al. 2022; Papanikolaou and Schmidt 2022). We show that remote work can also serve as a deliberate strategic lever through which startups can access and consequently reshape talent composition within firms. This perspective further extends prior research that put emphasis on performance outcomes, such as productivity, innovation, and job satisfaction (Dutcher 2012; Choudhury et al. 2021; Emanuel and Harrington 2024; Choudhury et al. 2026), by revealing how remote work reshapes startups' talent composition and organizational structure. It highlights remote work as an active hiring mechanism to rebuild the organization rather than solely an operational or flexibility policy, informing the debate over remote work as a sustained management strategy. We also extend this literature into startups, where prior work has been comparatively scarce despite startups' distinctive resource constraints and coordination needs.

Second, we contribute to the skill complementarity literature by shifting attention from its consequences — wages, tenure, mobility, and career outcomes — to its organizational antecedents. While prior work has established that complementarity is a key source of value creation and innovation, relatively little is known about how firms come to assemble complementary teams in the first place. By identifying remote work as an organizational mechanism that helps startups convert access to a more diverse talent pool into genuine skill complementarity, we help explain how complementarity is built, not just how it pays off once it exists. Relatedly, by documenting that increased complementarity is associated with greater organizational hierarchy, we connect the skill complementarity literature to organizational design research, showing that the benefits of complementary talent come bundled with coordination costs that firms address through structural adjustment.

Third, we contribute to the startup literature by identifying remote work as an IT-enabled hiring strategy through which startups reshape their talent composition. Prior research has shown that startup talent acquisition is influenced by founders' prior experiences, workers' preferences, and information frictions (e.g., Rocha and Grilli 2024; Roach and Sauermann 2024). A related stream documents how the resulting composition of startup human capital affects organizational development (e.g., Choi et al. 2025; Rocha and Brymer 2025). However, these streams of literature offer limited insights into how an IT-enabled work arrangement reconfigures the talent composition inside the firm. Although studies have shown that that remote work has an impact on startup labor markets (Hsu and Tambe 2025), whether this expanded applicant pool translates into realized hire that alter firms' internal skill composition remains unclear. We address this gap by showing that remote work enables startups to recruit workers whose skills both match their hiring needs and diversify the skill base of incumbent workers, thereby increasing workforce skill complementarity. We further show that this reconfiguration is associated with less flat organizational structures, revealing how remote hiring shapes not only the talent startups acquire but also how they organize and coordinate that talent.

**7.3. Managerial Implications**

Our findings also offer important managerial implications for startups. First, our study suggests that startups should view remote work as a deliberate hiring strategy, rather than simply as a workplace arrangement. Because early-stage firms operate under resource constraints and each hiring decision is crucial and each hiring decision can meaningfully shape the firm's trajectory (e.g., Stinchcombe 1965; Baker and Nelson 2005), startups should use remote hiring purposefully to improve team-level complementarity, as it relaxes the geographic and financial constraints that otherwise limit their access to talent. This is particularly true when startups can offer fully remote (rather than hybrid) arrangements and actively seek out workers whose skills are distinct (rather than duplicate).

Second, startups should recognize that the benefits of more complementary teams may come with greater coordination demands, leading to a more hierarchical structure. As startups adopt remote work and recruit workers with more complementary skills, they may need to introduce more formalized coordination mechanisms such as additional hierarchical structures to effectively integrate increasingly interdependent work. Consequently, these structural adjustments toward greater hierarchy may not necessarily be viewed as inefficiencies. Rather, they may represent a productive organizational response that enables startups to capture the value of a more complementary talent base, improve coordination, and build a stronger foundation for growth and execution.

### 7.4. Limitations and Future Research

This study has several limitations that open promising avenues for future research. First, our setting focuses on startups, which is theoretically important but may limit the generalizability of the results to larger and more established firms with more developed coordination systems. Future research could examine whether the relationship between remote work, skill complementarity, and hierarchy holds, strengthens, or weakens as firms mature and scale.

Second, although we show that remote work leads to greater hierarchy through skill complementarity, hierarchy reflects only one possible organizational response to a more complementary and diversified workforce. Startups may also adapt through other organizational changes, such as redesigning communication routines, relying more on informal coordination mechanisms, or adopting new collaboration technologies, which are not observable in this study. Future work could explore this broader set of organizational adaptations and examine whether they substitute for or complement the formal hierarchical adjustments we document in this study.

## Tables & Figures

**Figure 1 (a) Measuring Synergy**

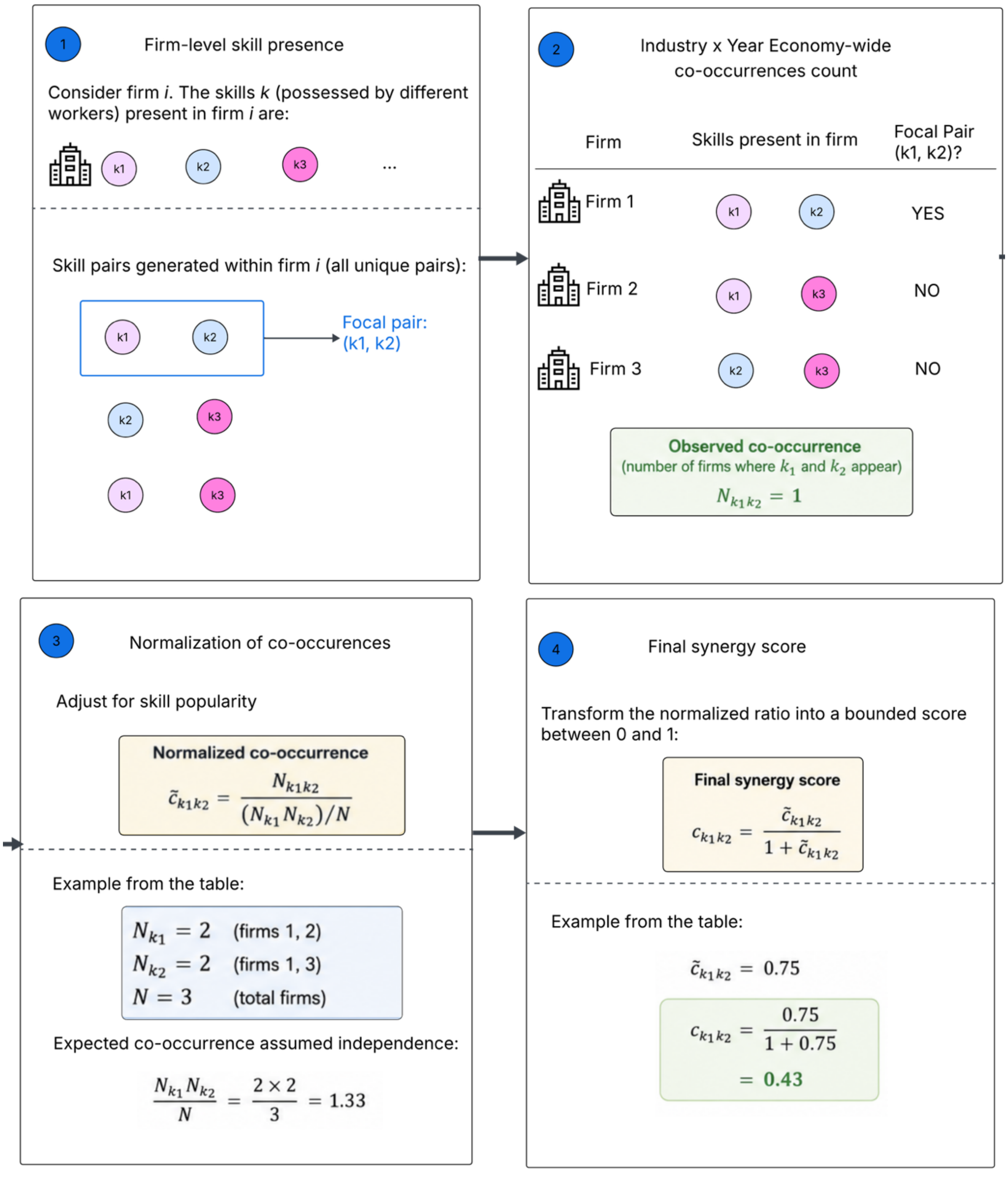

Notes: This figure illustrates how we construct the synergy score for each skill pair. For each firm $i$, we enumerate firm-level skill pairings—i.e., pairs of skills held by different workers within the same firm (to avoid "super-worker" co-occurrence). We then count how often each skill pair co-occurs across firms within the industry × year sector. Next, we compute the corresponding synergy measure $\widetilde{c_{k1k2}}$. Finally, we apply a monotone transformation to $c_{k1k2}$ to reduce distributional skew and place scores on a bounded scale. Note that $N_{k1}$ and $N_{k2}$ represent the number of firms with skill *k1* and *k2* presented, respectively. N represents the total number of firms in the economy. $N_{k1k2}$ is the number of firms where skill *k1* and *k2* both appear.

**Figure 1 (b) Measuring Substitutability**

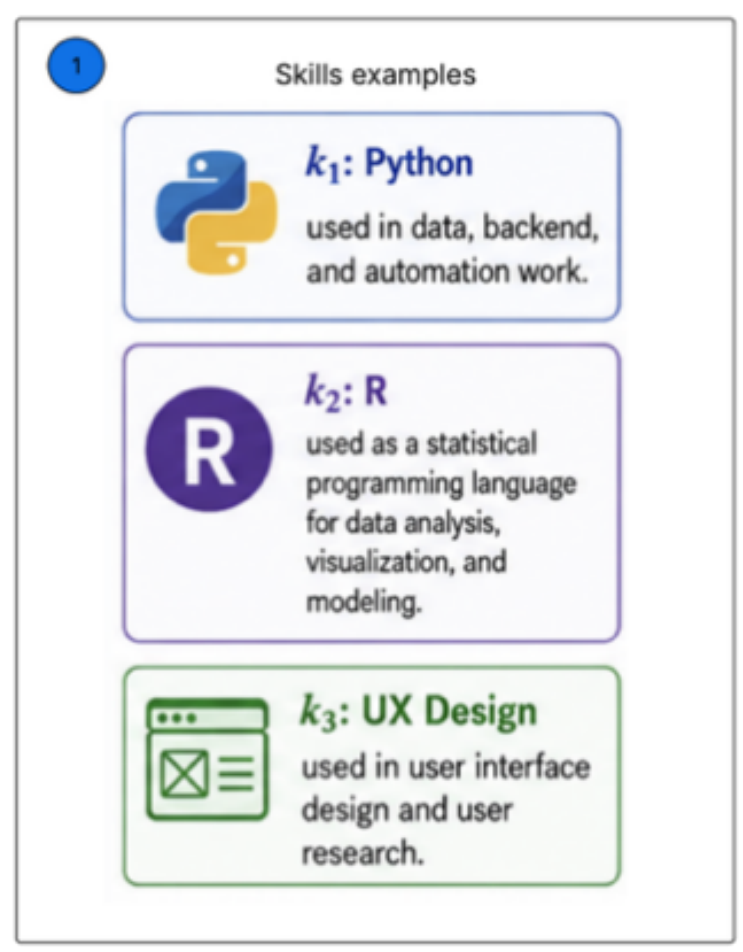

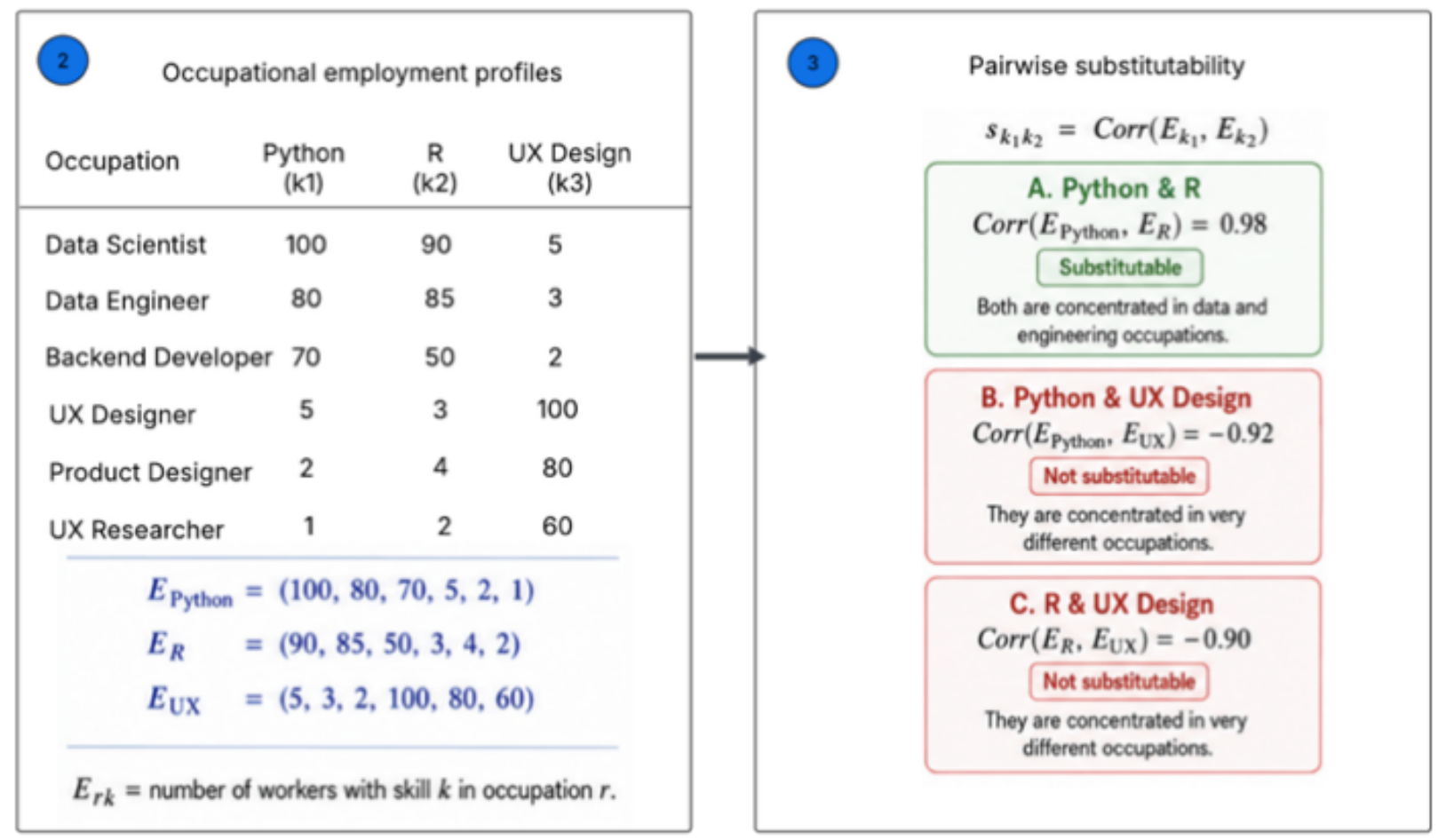

Notes: This figure illustrates how we construct the substitutability score for each skill pair. Workers with the skills *k1* and *k2* might choose different job roles. The substitutability between these skills is defined as the correlation between their occupational profiles.

**Figure 2. Worker-Level Complementarity Score Distribution**

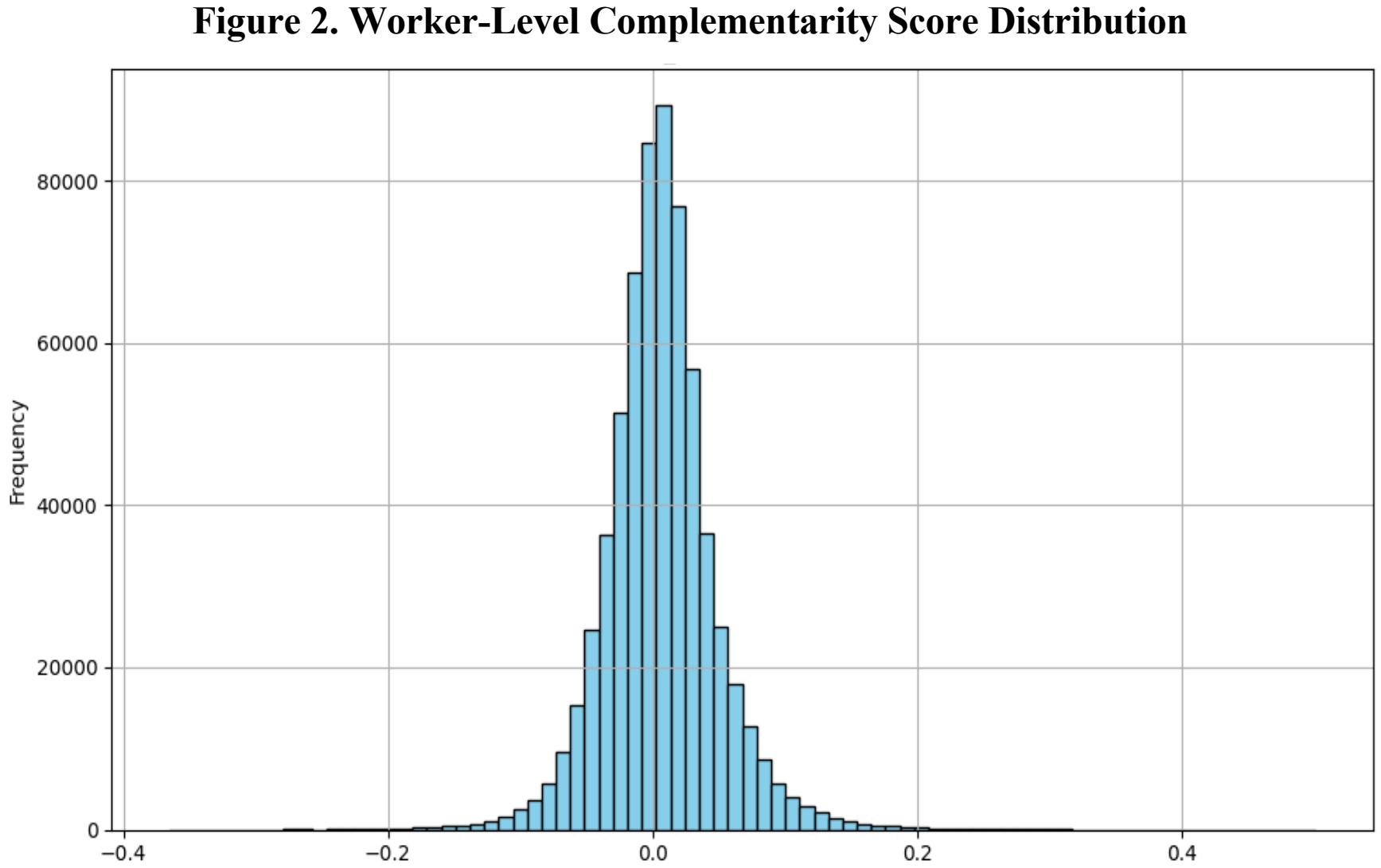

**Table 1. Summary Statistics**

| Variable Names | Obs | Mean | Min | Max | SD |
|---|---|---|---|---|---|
| Skill Complementarity | 29586 | 0.004 | -0.344 | 0.309 | 0.046 |
| Skill Complementarity (Alternative Measure) | 29586 | 0.456 | 0.042 | 0.857 | 0.066 |
| Total Number of Job Postings | 29586 | 3.584 | 0 | 8361 | 82.489 |
| Total Number of Workers | 29586 | 10.524 | 2 | 50 | 9.077 |
| Average Worker Annual Salary | 29586 | 94.023 | 42.24 | 685.81 | 67.66 |
| Organizational Flatness | 29586 | 9.455 | 1 | 50 | 8.256 |
| Remote Intensity | 29586 | 0.129 | 0 | 1 | 0.268 |
| Treatment | 29586 | 0.496 | 0 | 1 | 0.500 |

Notes: Summary statistics for the full sample period are reported at the firm-quarter level (29,586 observations). Salary values are scaled by 1,000; for example, a reported mean of 94.023 corresponds to an average annual salary of $94,023.

**Table 2. Impact of Remote Work on Workforce Skill Complementarity**

| | Skill_Complementarity | | | |
|---|---|---|---|---|
| | (1) | (2) | (3) | (4) |
| Remote_Intensity | 0.002** | 0.002** | 0.002* | 0.002** |
| | (0.001) | (0.001) | (0.001) | (0.001) |
| Quarterly FE | YES | YES | YES | YES |
| Firm FE | YES | YES | YES | YES |
| Controls | NO | YES | NO | YES |
| Observations | 29586 | 29586 | 27247 | 27247 |
| Number of Firms | 2286 | 2286 | 2284 | 2284 |

Notes: Robust standard errors are reported in parentheses. Columns (3) and (4) are based on lagged measure of remote work intensity. *** p < 0.01, ** p < 0.05, * p < 0.1.

**Table 3: 2SLS Estimation of Remote Work on Workforce Skill Complementarity**

| | Remote_Intensity (1) | | Skill_Complementarity (2) |
|---|---|---|---|
| Peer_Remote_Intensity | 0.073*** | Predicted_Remote_Intensity | 0.169* |
| | (0.006) | | (0.093) |
| Quarterly FE | NO | Quarterly FE | YES |
| Firm FE | NO | Firm FE | YES |
| Controls | YES | Controls | YES |
| Observations | 17205 | Observations | 17205 |
| Number of firms | 1309 | Number of firms | 1309 |

Notes: The table presents 2SLS estimates in which we use the remote work intensity of peer firms that are founded in similar years and located in the MSA but distant from the focal firm's industry as the instrumental variable. *** p < 0.01, ** p < 0.05, * p < 0.1.

**Table 4. How Remote Work Affects Organizational Flatness Through Workforce Skill Complementarity**

| | Organizational Flatness | | | | | |
|---|---|---|---|---|---|---|
| | (1) | (2) | (3) | (4) | (5) | (6) |
| Remote_Intensity | -0.408*** | | -0.400*** | -0.273** | | -0.269** |
| | (0.106) | | (0.145) | (0.114) | | (0.115) |
| Skill_Complementarity | | -3.837*** | -3.800*** | | -2.489** | -2.467** |
| | | (1.262) | (1.020) | | (1.107) | (1.104) |
| Quarterly FE | YES | YES | YES | YES | YES | YES |
| Firm FE | YES | YES | YES | YES | YES | YES |
| Controls | YES | YES | YES | YES | YES | YES |
| Observations | 29586 | 29586 | 29586 | 27247 | 27247 | 27247 |
| Number of Firms | 2286 | 2286 | 2286 | 2284 | 2284 | 2284 |

Notes: Organizational flatness is measured based on role taxonomy classifications. Columns (4) to (6) are based on lagged specifications, where organizational flatness is measured in the subsequent period. All specifications include firm and quarter fixed effects and control variables. *** p < 0.01, ** p < 0.05, * p < 0.1.

**Table 5: How Remote Work Increases Workforce Skill Complementarity**

| | Skill_Fit_Diversification | | Skill_Complementarity | |
|---|---|---|---|---|
| | (1) | | (2) | (3) |
| | | Skill_Fit_Diversification | 0.002*** | 0.002** |
| Remote_Intensity | 0.202*** | | (0.001) | (0.001) |
| | (0.020) | Remote_Intensity | | 0.003** |
| | | | | (0.001) |
| Quarterly FE | YES | Quarterly FE | YES | YES |
| Firm FE | YES | Firm FE | YES | YES |
| Controls | YES | Controls | YES | YES |
| Observations | 15363 | Observations | 15363 | 15363 |
| Number of Firms | 2175 | Number of Firms | 2175 | 2175 |

Notes: This table reports the mechanism analysis examining how remote work affects skill complementarity through greater skill fit and diversification among new hires. All models include firm and quarterly fixed effects, and the same set of control variables used in the baseline specification. Observations from firm-quarters without new hires are excluded because the measure is undefined when no incoming workers are present. As a result, the number of observations in this mechanism table is smaller than in the baseline analyses. *** $p < 0.01$, ** $p < 0.05$, * $p < 0.1$.

**Table 6. Effects of Remote Work Types (Fully vs. Hybrid) and Skill Similarity (Low vs. High) on Complementarity**

| | Skill_Complementarity | | | | |
|---|---|---|---|---|---|
| | (1) | (2) | (3) | (4) | (5) |
| **Panel A** | | | | | |
| FullRemoteIntensity | 0.003*** | | | | |
| | (0.001) | | | | |
| HybridRemoteIntensity | -0.005 | | | | |
| | (0.004) | | | | |
| **Panel B** | | | | | |
| HighSimilarity_ | | 0.001 | 0.001 | | |
| RemoteIntensity | | (0.002) | (0.003) | | |
| LowSimilarity_ | | 0.004** | 0.003*** | | |
| RemoteIntensity | | (0.001) | (0.001) | | |
| **Panel C** | | | | | |
| HighSimilarity_ | | | | 0.002 | 0.001 |
| FullRemoteIntensity | | | | (0.002) | (0.003) |
| LowSimilarity_ | | | | 0.005*** | 0.004*** |
| FullRemoteIntensity | | | | (0.002) | (0.001) |
| HighSimilarity_ | | | | -0.002 | -0.006 |
| HybridRemoteIntensity | | | | (0.006) | (0.004) |
| LowSimilarity_ | | | | -0.006 | -0.003 |
| HybridRemoteIntensity | | | | (0.005) | (0.007) |
| Quarterly FE | YES | YES | YES | YES | YES |
| Firm FE | YES | YES | YES | YES | YES |
| Controls | YES | YES | YES | YES | YES |
| Observations | 29586 | 29586 | 29586 | 29586 | 29586 |
| Number of Firms | 2286 | 2286 | 2286 | 2286 | 2286 |

Notes: Panel A decomposes remote intensity by work arrangement into fully remote and hybrid components. Panel B decomposes overall remote intensity into high- and low- skill similarity, where similarity captures the alignment between remote work job postings' required skills and the firm's existing workforce. Panel C jointly decomposes remote intensity by work arrangement and skill similarity. Columns (2) and (4) use the 50th percentile cutoff to define high vs. low similarity, whereas columns (3) and (5) use the 75th percentile cutoff. *** $p < 0.01$, ** $p < 0.05$, * $p < 0.1$.

**References**

Adams-Prassl, A., Boneva, T., Golin, M., & Rauh, C. (2022). Work that can be done from home: Evidence on variation within and across occupations and industries. *Labour Economics*, 74, 102083

Agrawal, A., Horton, J., Lacetera, N., & Lyons, E. (2015). Digitization and the contract labor market: A research agenda. In A. Goldfarb, S. M. Greenstein, & C. E. Tucker (Eds.), *Economic analysis of the digital economy (pp. 219–250).* University of Chicago Press.

Akan, M., Barrero, J. M., Bloom, N., Bowen, T., Buckman, S. R., Davis, S. J., & Kim, H. (2025). The new geography of labor markets (NBER Working Paper No. 33582). *National Bureau of Economic Research*. doi:10.3386/w33582

Aksoy, C. G., Barrero, J. M., Bloom, N., Davis, S. J., Dolls, M., & Zarate, P. (2022). Working from home around the world (NBER Working Paper No. 30446). *National Bureau of Economic Research.*

Alabdulkareem, A., Frank, M. R., Sun, L., AlShebli, B., Hidalgo, C. A., & Rahwan, I. (2018). Unpacking the polarization of workplace skills. *Science Advances*, 4(7), eaao6030.

Almeida, P., & Kogut, B. (1999). Localization of knowledge and the mobility of engineers in regional networks. *Management Science*, 45(7), 905–917.

Anderson, K. A. (2017). Skill networks and measures of complex human capital. *Proceedings of the National Academy of Sciences of the United States of America*, 114(48), 12720–12724.

Angelici, M., & Profeta, P. (2024). Smart working: Work flexibility without constraints. *Management Science*, 70(3), 1680–1705. doi:10.1287/mnsc.2023.4767

Arora, A., & Gambardella, A. (1990). "Complementarity and External Linkages: The Strategies of the Large Firms in Biotechnology*," Journal of Industrial Economics (38:4),* pp.361-379.

Bai, J. J., Brynjolfsson, E., Jin, W., Steffen, S., & Wan, C. (2021). Digital resilience: How work-from-home feasibility affects firm performance. NBER Working Paper No. 28588. *National Bureau of Economic Research*

Baker, T., & Nelson, R. E. (2005). Creating something from nothing: Resource construction through entrepreneurial bricolage. *Administrative Science Quarterly*, 50(3), 329–366.

Baron, R. M., & Kenny, D. A. (1986). The moderator-mediator variable distinction in social psychological research: Conceptual, strategic, and statistical considerations. *Journal of Personality and Social Psychology, 51, 1173-1182.*

Basu, S., Ma, X., & Shen, M. (2024). The value of human capital for firm performance: Roles of individual and group expertise. *Working paper.*

Becker, G. S., & Murphy, K. M. (1992). The division of labor, coordination costs, and knowledge. *The Quarterly Journal of Economics*, 107(4), 1137–1160. https://doi.org/10.2307/2118383

Beckman, C. M., Burton, M. D., & O'Reilly, C. (2007). Early teams: The impact of team demography on VC financing and going public. *Journal of Business Venturing*, 22(2), 147–173.

Bessen, J., Poege, F., & Röttger, R. (2023). Competing for talent: Large firms and startup growth. *Boston University School of Law.* Working paper.

Bloom, N., Liang, J., Roberts, J., and Ying, Z. J. (2015). Does working from home work? evidence from a chinese experiment. *The Quarterly Journal of Economics,* 130(1):165–218.

Boh, W., Constantinides, P., Padmanabhan, B., & Viswanathan, S. (2023). Building digital resilience against major shocks. *MIS quarterly,* 47(1), 343-360.

Bradley, S. W., Shepherd, D. A., & Wiklund, J. (2011). The importance of slack for new organizations facing "tough" environments. *Journal of Management Studies,* 48(5), 1071–1097.

Brattström, A. (2024). Task re-allocation in new venture teams: A team conflict perspective. *Entrepreneurship Theory and Practice,* 48(1), 205-245.

Brynjolfsson, E., & Milgrom, P. (2013). "Complementarity in Organizations," in The Handbook of Organizational Economics, R. Gibbons and J. Roberts (eds.), Princeton*, NJ: Princeton University Press*, pp. 11-55.

Burton, M. D., Dahl, M. S., & Sorenson, O. (2018). Do start-ups pay less? *ILR Review*, 71(5), 1179–1200.

Callaway, B., & Sant'Anna, P. H. C. (2021). Difference-in-differences with multiple time periods. *Journal of Econometrics,* 225(2), 200–230.

Campbell, B. A., Coff, R., & Kryscynski, D. (2012). Rethinking sustained competitive advantage from human capital. *Academy of Management Review,* 37(3), 376–395.

Castellaneta, F., Conti, R., & Kacperczyk, A. J. (2026). Gender gap in startup recruiting: Evidence from changes in termination costs. *Management Science*, 72(4), 3571–3591. doi:10.1287/mnsc.2023.01491

Choi, J., Goldschlag, N., Haltiwanger, J. C., & Kim, J. D. (2025). Early joiners and startup performance. *The Review of Economics and Statistics,* 107*(6), 1485–1500.* doi:10.1162/rest_a_01386

Choudhury, P., Foroughi, C., and Larson, B. (2021). Work-from-anywhere: The productivity effects of geographic flexibility. *Strategic Management Journal*, 42(4):655–683.

Choudhury, P., Khanna, T., Makridis, C. A., and Schirmann, K. (2026). Is hybrid work the best of both worlds? evidence from a field experiment. *The Review of Economics and Statistics, 108*(4), 1134–1140.

Cohen, S. L., Bingham, C. B., & Hallen, B. L. (2019). The Role of Accelerator Designs in Mitigating Bounded Rationality in New Ventures. *Administrative Science Quarterly,* 64(4), 810-854.

Cohen, W. M., & Levinthal, D. A. (1990). Absorptive capacity: A new perspective on learning and innovation. *Administrative Science Quarterly,* 35(1), 128–152.

Danaher, B., Hersh, J., Smith, M. D., & Telang, R. (2020). The effect of piracy website blocking on consumer behavior. *MIS Quarterly, 44*(2), 631–659.

Dennis, A. R., Fuller, R. M., & Valacich, J. S. (2008). Media, tasks, and communication processes: A theory of media synchronicity. *MIS Quarterly*, 32(3), 575–600.

Dettling, L. J. (2017). Broadband in the labor market: The impact of residential high-speed internet on married women's labor force participation. *Industrial and Labor Relations Review*, 70(2), 451–482.

DiMaggio, P. J., & Powell, W. W. (1983). The iron cage revisited: Institutional isomorphism and collective rationality in organizational fields*. American Sociological Review,* 48(2), 147–160.

Dutcher, E. G. (2012). The effects of telecommuting on productivity: An experimental examination. the role of dull and creative tasks. *Journal of Economic Behavior & Organization*, 84(1):355–363.

Eisenhardt, K. M., & Schoonhoven, C. B. (1990). Organizational growth: Linking founding team, strategy, environment, and growth among U.S. semiconductor ventures, 1978–1988. *Administrative Science Quarterly*, 35(3), 504–529.

Ellison, N. B. (1999). Social Impacts: New Perspectives on Telework. *Social Science Computer Review*, 17(3), 338-356.

Emanuel, N. and Harrington, E. (2024). Working remotely? selection, treatment, and the market for remote work. *American Economic Journal: Applied Economics*, *16(4),* 528-559.

Ennen, E., & Richter, A. (2010). The whole is more than the sum of its parts—Or is it? A review of the empirical literature on complementarities in organizations. *Journal of Management,* 36(1), 207–233.

Garicano, L. (2000). Hierarchies and the organization of knowledge in production. *Journal of Political Economy*, 108(5), 874–904.

Gibbs, M., Mengel, F., and Siemroth, C. (2023). Work from home and productivity: Evidence from personnel and analytics data on information technology professionals*. Journal of Political Economy Microeconomics*, 1(1):7–41.

Grant, R. M. (1996). Toward a knowledge-based theory of the firm. *Strategic Management Journal*, 17(Winter Special Issue), 109–122.

Guzman, J. (2024). Go west young firm: The impact of startup migration on the performance of migrants. *Management Science, 70(7)*, 4824–4846.

Hellmann, T., & Puri, M. (2002). Venture capital and the professionalization of start-up firms: Empirical evidence. *The Journal of Finance*, 57(1), 169–197.

Hou, J., Liang, C., Chen, P.-Y., & Gu, B. (2024). Can telework adjustment help reduce disaster-induced gender inequality in job market outcomes? *Information Systems Research*, 35(4), 1701–1720.

Hsu, D. H., & Tambe, P. B. (2025). Remote work and job applicant diversity: Evidence from technology startups. *Management Science*, 71(1), 595–614. doi:10.1287/mnsc.2022.03391

Jaffe, A. B., Trajtenberg, M., & Henderson, R. (1993). Geographic localization of knowledge spillovers as evidenced by patent citations. *Quarterly Journal of Economics*, 108(3), 577–598.

Kent, S. L. (2001).The ultimate history of video games: From pong to Pokemon—The story behind the craze that touched our lives and changed the world. *New York, NY: Crown.*

Kim, J. D. (2018). Is there a startup wage premium? Evidence from MIT graduates. *Research Policy,* 47(3), 637–649

Krusell, P., Ohanian, L. E., Ríos-Rull, J.-V., & Violante, G. L. (2000). Capital-skill complementarity and inequality: A macroeconomic analysis. *Econometrica,* 68(5), 1029–1053.

Lawrence, M., & Poliquin, C. (2023), The growth of hierarchy in organizations: Managing knowledge scope, *Strategic Management Journal,* 44(13), 3155–3184.

Lee, S. R. (2021). The myth of the flat start-up: Reconsidering the organizational structure of start-ups. *Strategic Management* Journal, 43(1), 58–92. doi:10.1002/smj.3333

Lee, S. Y., Florida, R., & Acs, Z. (2004). Creativity and Entrepreneurship: A Regional Analysis of New Firm Formation. *Regional Studies*, 38, 879–891.

Lindquist, M. J. (2004). Capital-skill complementarity and inequality over the business cycle. *Review of Economic Dynamics,* 7(3), 519–540.

Mas, A., & Pallais, A. (2017). Valuing alternative work arrangements. *American Economic Review,* 107(12), 3722–3759.

McEvily, B., Soda, G., & Tortoriello, M. (2014). More formally: Rediscovering the missing link between formal organization and informal social structure. *Academy of Management Annals,* 8*(1), 299–345.*

Meier, S., Stephenson, M., & Perkowski, P. (2019). Culture of trust and division of labor in nonhierarchical teams. *Strategic Management Journal*, 40(8), 1171–1193

Milgrom, P., & Roberts, J. (1995). Complementarities and fit strategy, structure, and organizational change in manufacturing. *Journal of Accounting and Economics*, 19(2–3), 179–208.

Moscelli, G., Sayli, M., Mello, M., & Vesperoni, A. (2025). Staff engagement, co-workers' complementarity and employee retention: Evidence from English NHS hospitals. *Economica,* 92(365), 42–83. doi:10.1111/ecca.12554

Nath, S., Yeo, J., Bharadwaj, A., & Bharadwaj, S. G. (2026). Pricing of multidimensional skills in IT labor market: Complementarities between AI and traditional IT skills [Preprint]. *SSRN*. doi:10.2139/ssrn.6188398

Nedelkoska, L., & Neffke, F. (2019). Skill mismatch and skill transferability: Review of concepts and measurements. *Papers in Evolutionary Economic Geography,* 19(21).

Neffke, F. M. H. (2019). The value of complementary co-workers. *Science Advances, 5*(12), eaax3370.

Neffke, F., Otto, A., & Weyh, A. (2017). Inter-industry labor flows. *Journal of Economic Behavior & Organization,* 142*, 275–292.*

Olson, M. H. (1983). Remote office work: Changing work patterns in space and time. *Communications of the ACM, 26*(3), 182–187.

Papanikolaou, D., & Schmidt, L. D. (2022). Working remotely and the supply-side impact of Covid-19. *The Review of Asset Pricing Studies*, 12(1), 53–111.

Ployhart, R. E., & Moliterno, T. P. (2011). Emergence of the human capital resource: A multilevel model. *Academy of Management Review,* 36(1), 127–150

Ployhart, R. E., Nyberg, A. J., Reilly, G., & Maltarich, M. A. (2014). Human capital is dead; long live human capital resources! *Journal of Management, 40*(2), 371–398.

Preacher, K. J., & Hayes, A. F. (2008). Asymptotic and resampling strategies for assessing and comparing indirect effects in multiple mediator models. *Behavior Research Methods, 40*(3), 879–891.

Puranam, P., Raveendran, M., & Knudsen, T. (2012). Organization design: The epistemic interdependence perspective. *Academy of Management Review*, 37(3), 419–440.

Puri, M., & Zarutskie, R. (2012). On the life cycle dynamics of venture-capital- and non-venture-capital-financed firms. *The Journal of Finance, 67(6), 2247–2293*

Roach, M., & Sauermann, H. (2024). Can technology startups hire talented early employees? Ability, preferences, and employee first job choice. *Management Science,* 70(6), 3619–3644.

Rocha, V., & Brymer, R. A. (2025). We go way back: Affiliation-based hiring and young firm performance. *Strategic Management Journal,* 46(3), 723–749.

Rocha V, Grilli L. (2024) Early-stage start-up hiring: The interplay between start-ups' initial resources and innovation orientation. *Small Bus. Econom*. 62(4):1641–1668.

Shaver, J. M., & Flyer, F. (2000). Agglomeration economies, firm heterogeneity, and foreign direct investment in the United States. *Strategic Management Journal, 21*(12), 1175–1193.

Sine, W. D., Mitsuhashi, H., & Kirsch, D. A. (2006). Revisiting Burns and Stalker: Formal structure and new venture performance in emerging economic sectors*. Academy of Management Journal, 49*(1), 121–132.

Song, J., Almeida, P., & Wu, G. (2003). Learning-by-hiring: When is mobility more likely to facilitate interfirm knowledge transfer? *Management Science*, 49(4), 351–365.

Sorenson, O., Dahl, M. S., Canales, R., & Burton, M. D. (2021). Do startup employees earn more in the long run? *Organization Science, 32*(3), 587–604

Staples, D. S., & Webster, J. (2008). Exploring the effects of trust, task interdependence and virtualness on knowledge sharing in teams. *Information Systems Journal,* 18(6), 617–640.

Stephany, F., & Teutloff, O. (2024). What is the price of a skill? The value of complementarity. *Research Policy*, 53(1), 104898.

Stinchcombe, A. L. (1965). Social structure and organizations. In J. G. March (Ed.), *Handbook of organizations* (pp. 142–193). Rand McNally.

Torres, C. I., & Crossler, R. E. (2025). Promoting security behaviors in remote work environments: Personal values shaping information security policy compliance. *Information Systems Research*, *36*(2), 1183-1195.

Witman, A., Beadles, C., Liu, Y., Larsen, A., Kafali, N., Gandhi, S., Amico, P., & Hoerger, T. (2019). Comparison group selection in the presence of rolling entry for health services research: Rolling entry matching. *Health Services Research*, 54(2), 492–501.

Wright, P. M., Coff, R., & Moliterno, T. P. (2014). Strategic human capital: Crossing the great divide. *Journal of Management*, 40(2), 353–370.

Yang, L., Holtz, D., Jaffe, S. et al. The effects of remote work on collaboration among information workers. *Nat Hum Behav* 6, 43–54 (2022). https://doi.org/10.1038/s41562-021-01196-4

## Online Appendix

### A. Balance checks

After applying rolling entry matching (REM), our final sample includes 2,286 startups, comprising 1,143 startups that adopted remote work at some point after they were founded and 1,143 matched startups that did not adopt remote work since the founding date until the end of our sample period. Table A below reports the balance check result. The two groups are well balanced on observable pre-treatment characteristics, with no statistically significant differences between the treatment and control groups, confirming that the matched sample achieves good covariate balance prior to the shock.

Table A. Balance Check

| Variable Names | Obs | Treatment | Control | Difference | P-value |
|---|---|---|---|---|---|
| Skill Complementarity | 2286 | 0.007 | 0.005 | 0.002 | 0.171 |
| Skill Complementarity (Alternative Measure) | 2286 | 0.459 | 0.457 | 0.002 | 0.482 |
| Total Number of Job Postings | 2286 | 1.199 | 0.996 | 0.203 | 0.277 |
| Total Number of Workers | 2286 | 7.090 | 6.908 | 0.182 | 0.441 |
| Average Worker Annual Salary | 2286 | 106.37 | 102.19 | 4.180 | 0.174 |
| Organizational Flatness | 2286 | 6.556 | 6.292 | 0.264 | 0.225 |

Notes: Balance checks are computed in the pre-treatment period at the firm level (2,286 matched treatment–control pairs). Salary values are scaled by 1,000. The rightmost columns report mean differences between treatment and control firms prior to the shock and the corresponding p-values from two-sample t-tests, confirming that covariates are balanced after 1–1 matching.

## B. Extension of Neffke (2019)

To measure the skill complementarity of a startup's workforce, we advance Neffke 2019's method in three ways. First, we shift the unit of analysis from educational tracks to fine-grained skill-level data provided by the Revelio Lab. Specifically, Neffke 2019's approach was originally designed based on educational backgrounds, which are relatively coarse in capturing workers' capabilities. In contrast, our approach is able to measure complementarity at the level of fine-grained skill attributes rather than broad educational categories, enabling a more precise assessment of how workers contribute complementary capabilities to their teams.

Second, we refine the construction of co-occurrence by focusing on cross-worker co-occurrence of skill pairs. The skill in Neffke 2019 is based on educational tracks, and workers typically hold only one educational field, making within-person co-occurrence of skill pairs extremely rare in that setting. In the Revelio data, however, a single worker usually has multiple skills. As a result, skill pairs may appear together not because they represent cross-worker synergy, but because these skill pairs are possessed by the same person. Such cases can artificially inflate synergy scores and obscure the extent to which different workers contribute complementary capabilities. To better capture cross-worker synergy, we therefore refine the co-occurrence calculation by considering only skill pairs that appear across different workers within the same firm. This ensures that the resulting synergy measure captures team-based complementarities rather than within-person skill co-occurrence.

Third, we extend the framework by constructing segment-specific dictionaries of skill networks. The original approach relies on a single, economy-wide network of skill synergy and substitutability. However, skill networks naturally vary across industries and evolve over time. Using a single dictionary of skill relationships would obscure these important differences. To address this limitation, we group observations into year-industry combinations and construct segment-specific dictionaries of synergy and substitutability scores for startup firms. This stratification accounts for temporal and sectoral heterogeneity in skill patterns, ensuring that complementarity estimates reflect the unique dynamics within each industry and year for startups.

## C. Skill Complementarity Measurement

### C.1. Descriptive Results on Skill Complementarity

Tables C1 and C2 below provide illustrative examples of the synergy and substitutability scores for skill pairs of several industry sectors, highlighting the skill pairs with the highest and lowest synergy scores and those with the highest and lowest substitutability scores.

Table C1. Synergy Scores by Industry x Year Sector Sectors

| Industry | Max Synergy Score | Skill Pair | Min Synergy Score | Skill pair |
|---|---|---|---|---|
| Human Resources Services | 0.8648 | **Skill 1**: emergency management / physical security / crisis management<br><br>**Skill 2**: psychology / mental health / treatment | 0.2209 | **Skill 1**: architecture / aviation / aircraft<br><br>**Skill 2**: healthcare / hospitals / healthcare management |
| Healthcare and Wellness Services | 0.8852 | **Skill 1**: clinical research / clinical trials / medicine<br><br>**Skill 2**: medical devices / pharmaceutical sales / cardiology | 0.3421 | **Skill 1**: chemistry / molecular biology / biochemistry<br><br>**Skill 2**:emergency management / physical security / crisis management |

Notes: This table reports illustrative examples of synergy scores for skill pairs in two sectors in the year 2023. For each sector, we present the highest- and lowest-synergy pairs, along with their corresponding skill combinations. Within each skill, slash-separated terms (e.g., *medical devices/pharmaceutical sales/cardiology*) represent semantically related skill labels grouped into a single skill, whereas **Skill 1** and **Skill 2** together constitute the reported skill pair.

Table C2. Substitutability Scores by Industry x Year Sector Sectors

| Industry | Max Substitutability Score | Max Skill Pair | Min Substitutability Score | Min Skill pair |
|---|---|---|---|---|
| Information Technology Services | 0.7759 | **Skill 1:** html / javascript / mysql<br><br>**Skill 2**: java / eclipse / android | -0.0120 | **Skill 1**: linux / unix / perl<br><br>**Skill 2**: management / leadership / training |
| Healthcare and Wellness Services | 0.7451 | **Skill 1**: clinical research / clinical trials / medicine<br><br>**Skill 2**: healthcare / hospitals / healthcare management | -0.1343 | **Skill 1**:microsoft office / microsoft excel / microsoft word<br><br>**Skill 2**: testing / automation / electrical engineering |

Notes: This table reports illustrative examples of substitutability scores for skill pairs in two sectors in the year 2023. For each sector, we present the highest- and lowest- substitutability pairs, along with their corresponding skill combinations. Within each skill, slash-separated terms (e.g., *html/javascript /mysql)* represent semantically related skill labels grouped into a single skill, whereas **Skill 1** and **Skill 2** together constitute the reported skill pair.

Figure C1 below plots average skill synergy against substitutability by industry, with the vertical distance from the fitted line indicating average co-worker complementarity. Knowledge-intensive sectors, such as Information Technology Services, Biotech and Healthcare Services, Research and Development, and Financial Services, generally lie above the line, whereas more operational sectors, including Apparel Retail, Logistics and Transportation, Food and Beverage, and Retail and Consumer Goods, tend to lie below it. Creative sectors cluster closer to the baseline, suggesting a mix of complementary and more interchangeable roles.

Figure C1. Average co-worker synergy vs. substitutability by industry

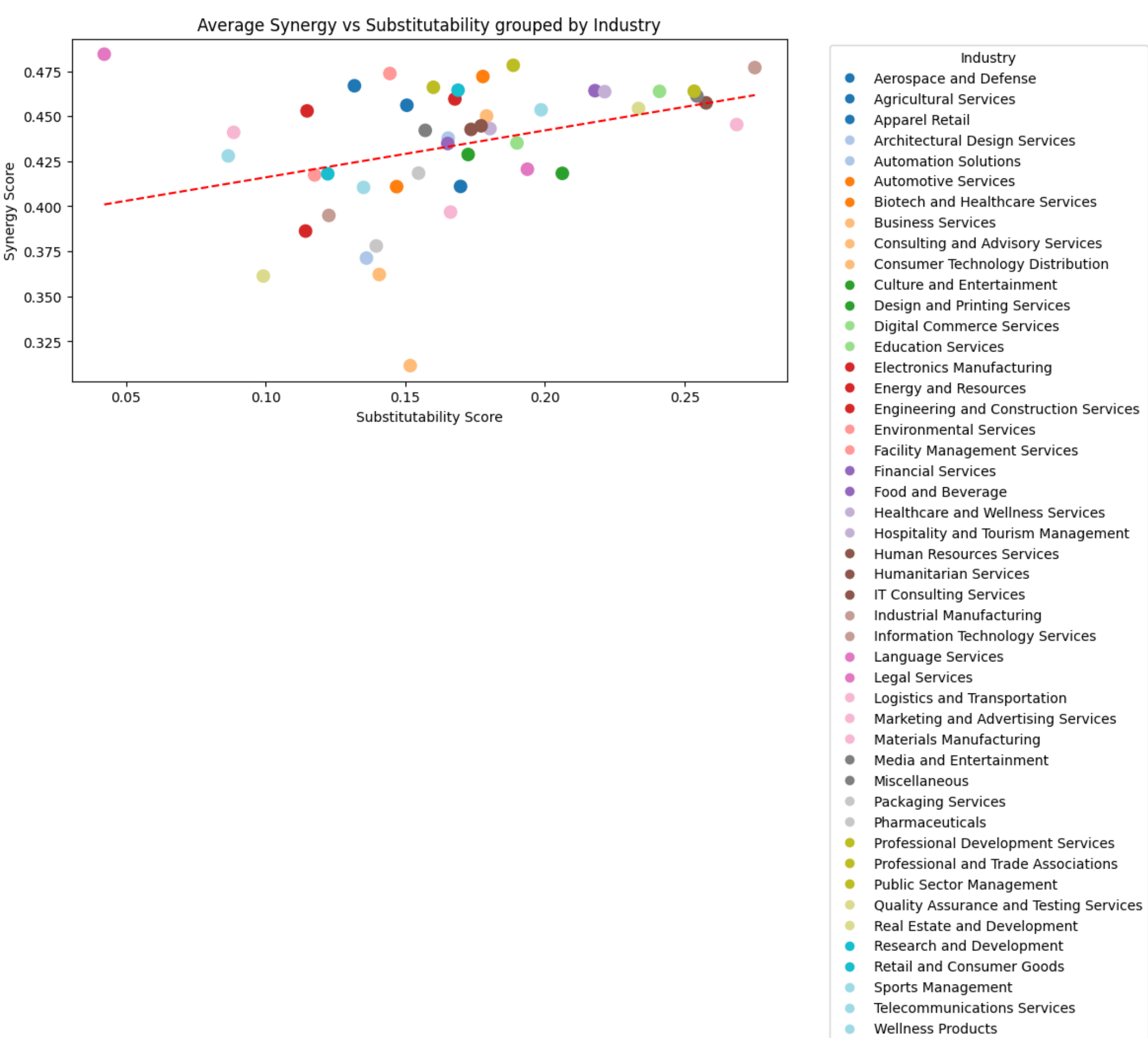

Notes: This figure plots the average substitutability and synergy scores across industries. Each point represents an industry. For each industry, we calculate the average substitutability score and average synergy score. The horizontal axis reports the industry-average substitutability score, and the vertical axis reports the industry-average synergy score. The dashed red line represents the fitted linear regression line. Industries above the line tend to hire a workforce with high co-worker skill complementarity, whereas industries below the line exhibit a more homogeneous workforce.

### C.2. Wage Prediction

To further validate our measure of workforce skill complementarity, we examine its relationship with average workforce wages. As shown in Table C3, workforce skill complementarity is positively and significantly associated with average wages. This finding is consistent with prior research documenting wage premiums associated with skill complementarities and provides additional support for the validity of our measure.

Table C3. Impact of Skill Complementarity on Average Salary

| | Average_Salary | |
|---|---|---|
| | (1) | (2) |
| Skill_Complementarity | 0.611*** | 0.776*** |
| | (0.139) | (0.146) |
| Quarterly FE | YES | YES |
| Firm FE | YES | YES |
| Controls | NO | YES |
| Observations | 29586 | 29586 |
| Number of Firms | 2286 | 2286 |

Notes: This table reports the relationship between workforce skill complementarity and average salary at the firm-quarter level. p < 0.01, ** p < 0.05, * p < 0.1

### C.3. Alternative Measure of Skill Complementarity

This section provides additional details on the alternative measure of workforce skill complementarity based on Alabdulkareem et al. (2018). Alabdulkareem et al. (2018) defines complementarity at the skill-pair level using O*NET data on occupation-skill importance scores. Their measure is constructed from the frequency with which two skills are jointly used within the same occupations, relative to their standalone occurrence rates. As a result, the Alabdulkareem et al.(2018) measure captures complementarities between skills, abstracted from firm-level employment networks. The framework proceeds in three steps: defining occupation–skill importance, identifying effectively used skills via revealed comparative advantage (RCA), and constructing pairwise skill complementarity.

To apply the method in our setting, we first construct an occupation–skill matrix by mapping our firm-level worker data to O*NET skill information. Specifically, we collect the $O$*NET occupation–skill importance scores $onet(j, s)$. We then link these O*NET occupation–skill scores to our workforce data in two steps. First, job roles in our dataset are directly matched to standardized $O$*NET occupation codes using the existing occupation mapping provided in the dataset. This yields a consistent occupation-level mapping for each worker. Second, skill categories in our dataset are aligned with O*NET skill definitions. Because naming conventions and skill descriptions may not perfectly coincide across datasets, we employ large language model (LLM) assistance to match skill labels and descriptions to the closest corresponding $O$*NET skill categories. This process ensures semantic consistency while preserving the underlying O*NET skill taxonomy.

Let $onet(j, s)$ denotes the importance or intensity of skill $s$ in occupation $j$, obtained from O*NET. These values range from 0 to 1 (or are normalized importance scores). The score 1 indicates that this skill $s$ is essential to $j$, while $onet(j, s) = 0$ indicates that workers of occupation $j$ need not possess or perform $s$.

Next, not all skills listed in O*NET are equally central to an occupation. To identify which skills are effectively used, we employ a Revealed Comparative Advantage (RCA) metric:

$$\mathrm{RCA}(j, s) = \frac{\mathrm{onet}(j,s) / \sum_{s' \in S} \mathrm{onet}(j,s')}{\left(\sum_{j' \in J} \mathrm{onet}(j',s)\right) / \left(\sum_{j' \in J} \sum_{s' \in S} \mathrm{onet}(j',s')\right)} \tag{1}$$

We then define the effective use of a skill within an occupation based on the revealed comparative advantage (RCA) threshold. Specifically, we construct a binary indicator:

$$e(j, s) = 1\{\mathrm{RCA}(j, s) > 1\} \tag{2}$$

The indicator $\mathrm{e(j, s)}$ equals one when skill $s$ is relatively overrepresented in occupation $j$ compared to its overall prevalence across all occupations. Intuitively, this step retains only those skills that are core to a given occupation.

Using this binary occupation–skill matrix, we measure complementarity between two skills $s$ and $s'$ based on their co-occurrence across occupations. It is defined by the minimum of the conditional probabilities of a pair of skills being effectively used by the same occupation:

$$\theta(s, s') = \frac{\sum_{j \in J} e(j,s) e(j,s')}{\max\left(\sum_{j \in J} e(j,s), \sum_{j \in J} e(j,s')\right)} \tag{3}$$

Higher values of $\theta(s, s')$ indicate that two skills are more frequently and effectively used together across occupations, reflecting stronger structural complementarity. Conversely, lower values suggest that the skills are less systematically bundled together and therefore exhibit weaker interdependence.

Building on this pairwise measure, we next construct worker-level skill complementarity. We then aggregate these worker-level complementarity measures to the firm–quarter level by taking the average across all workers within the firm *i* at time *t*. This aggregation yields a firm–quarter measure of workforce skill complementarity that reflects the overall degree to which skill bundles within the firm are mutually reinforcing. Figure C2 below compares the distribution of the two measurements.

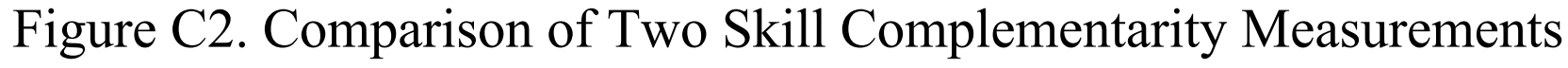

Figure C2. Comparison of Two Skill Complementarity Measurements

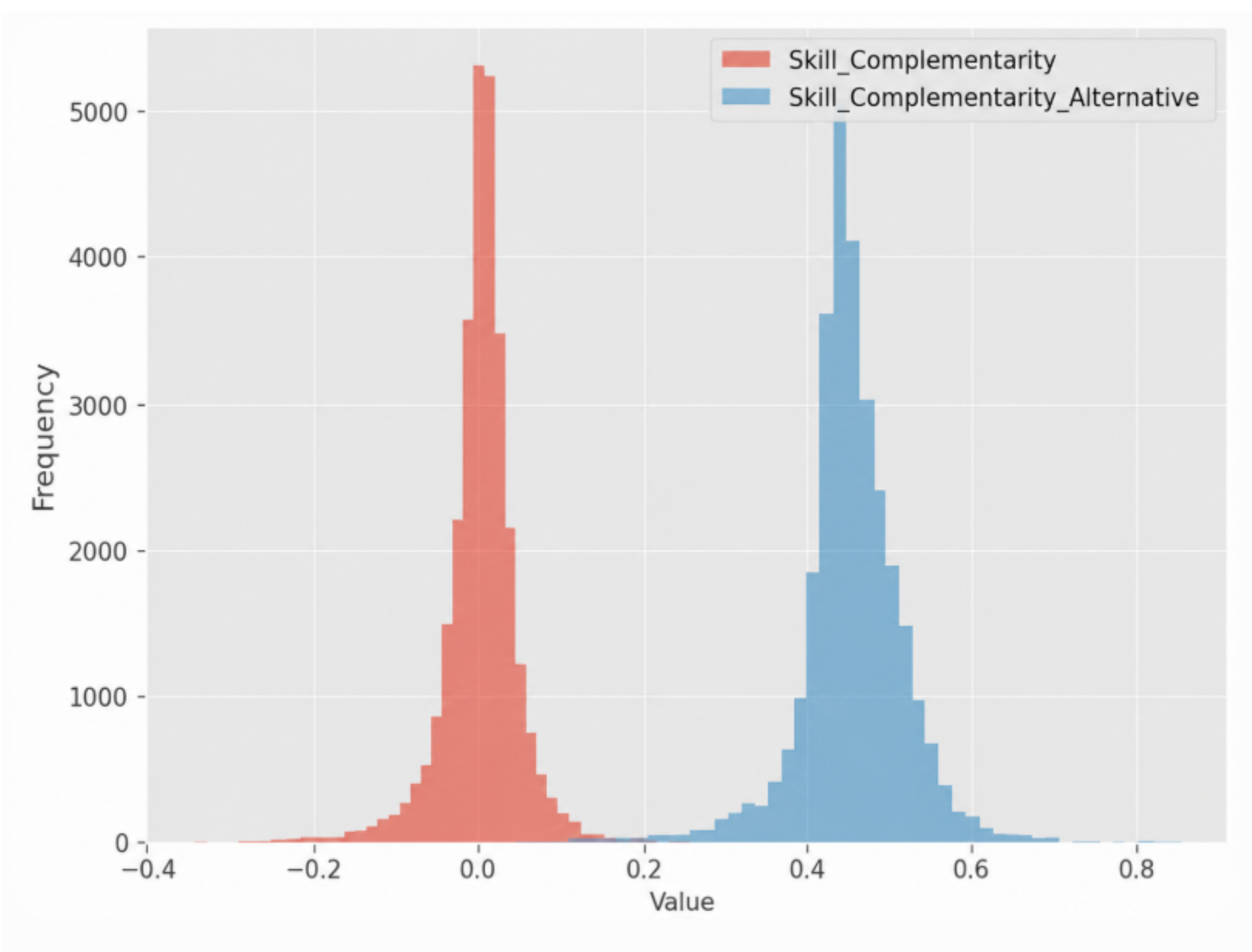

## D. Instrumental Variable Estimation

Our instrumental variable leverages variation in remote work adoption among firms that are geographically proximate to the focal firm but operate in industries with distinct human capital requirements. The reasons are as follows. Those firms located in the same metropolitan statistical area (MSA) and founded in similar years are likely to face similar local conditions that influence the adoption of remote work of the focal firm. At the same time, those firms operate in industries with low similarity in human capital requirements are unlikely to directly affect the focal firm's workforce.

To operationalize this measure, we proceed as follows. First, for each focal firm, we identify its headquarters location at the MSA level and its primary industry classification. Second, within each founded-year × MSA category, we identify peer firms that are founded in the close years and located in the same MSA but operate in industries different from that of the focal firm. Among these candidate peers, we restrict to firms in the most distant industries according to an industry-relatedness measure by Neffke et al. (2017), which captures low similarity in human capital requirements as inferred from limited inter-industry labor mobility.

Then, we gather these companies' job postings for our sample period 2021Q1 to 2025Q2. Finally, analogous to the construction of our dependent variable, we compute a cumulative remote intensity measure over time to capture the peer remote adoption. This cumulative peer remote intensity serves as the instrumental variable in our IV specification $Peer_Remote_Intensity_{it}$.

In our setting, the exclusion restriction assumption is satisfied because the peer firms are drawn from the most distant industries with low similarity in human capital requirements relative to the focal firm. The industry-relatedness measurement by Neffke et al. (2017) quantifies similarity in human capital requirements across industries. Their framework shows that industries with low relatedness exhibit low inter-industry labor mobility, reflecting limited overlap in skill demands and industry-specific human capital. For example, the most distant industry relative to Humanitarian Services is Electronics Manufacturing. Therefore, the peer firms used in our instrument operate in sectors with fundamentally different human capital structures, making it unlikely that their remote work adoption directly affects the focal firm's workforce skill composition through labor market spillovers. Instead, any influence of peer remote intensity on the focal firm's organizational outcomes operates primarily through its effect on the focal firm's own remote work adoption, thereby satisfying the exclusion restriction.

The relevance condition requires that the instrument meaningfully predicts the focal firm's remote work intensity. In our context, peer remote intensity provides exogenous variation because firms systematically adjust decision-making in response to peer behavior and broader managerial trends. A large body of research documents peer effects and policy diffusion in organizational choices driven by information spillovers (DiMaggio and Powell, 1983; Leary and Roberts 2014; Shue 2013). This mechanism is especially salient in the post-covid period, during which remote work evolved into a strategic organizational choice rather than a temporary operational response. Consequently, increases in remote work intensity among these peers influence the focal firm's likelihood of adopting or expanding its own remote work practices, satisfying the first-stage relevance requirement.

Overall, our instrument satisfies both the relevance and exogeneity conditions required for valid identification. Accordingly, we employ this instrument in a two-stage least squares (2SLS) framework to estimate the causal effect of remote work intensity on skill complementarity. To construct the instrument, we expand the founding-year bandwidth to ±3 years around each focal firm to increase the number of candidates for peer firm to select those with the least human capital similarity. We then measure *Peer_Remote_Intensity* as the remote adoption among these peer firms.

## E. Robustness checks

### E.1. The Impact of Remote work on Skill Complementarity (Alternative Measure)

Table E1 below shows the impact of remote work on this alternative measure of skill complementarity. The estimates remain positive and statistically significant with and without control variables, consistent with those obtained using the original measure.

Table E1. Impact of Remote Work on Workforce Skill Complementarity (Alternative Measure)

| | Skill_Complementarity_Alternative | |
|---|---|---|
| | (1) | (2) |
| Remote_Intensity | 0.004* | 0.004* |
| | (0.002) | (0.002) |
| Quarterly FE | YES | YES |
| Firm FE | YES | YES |
| Controls | NO | YES |
| Observations | 29586 | 29586 |
| Number of Firms | 2286 | 2286 |

Notes: The dependent variable, *Skill_Complementarity Alternative*, is defined as the average complementarity score across all workers within a start-up using alternative skill complementarity measure. *** $p < 0.01$, ** $p < 0.05$, * $p < 0.1$.

### E.2. Pre-trend Analysis

To further validate our difference-in-differences (DID) identification strategy, we conduct timing falsification tests designed to assess the plausibility of the parallel trend assumption. A key requirement for causal interpretation in a DID framework is that treatment and control firms would have followed similar pre-treatment trajectories. In our context, this implies that firms with remote work adoption should not exhibit systematically different trends in skill complementarity prior to remote work adoption.

To test this, we estimate event-time specifications that examine the dynamic relationship between remote intensity and workforce skill complementarity in periods leading up to remote adoption, where relative time represents the number of periods before or after a firm's initial remote adoption, with negative values indicating pre-treatment periods:

$$Skill_Complementarity_{it} = \alpha_t + \sum_{k=-3}^{-1} \gamma_k \left( Treatment_f \times \mathbf{1}\{relative_time_{it} = k\} \right) + \quad (4)$$

$$\beta 1 \cdot Remote_Intensity_{it} \; x \; Post + \gamma_t + \mu_i + \lambda X_{it} + \varepsilon_{it}$$

The regression results are reported in Table E2 below. The coefficients corresponding to the pre-treatment periods are statistically insignificant. This pattern is consistent with the parallel trends assumption, indicating that control and treated firms do not exhibit systematically different trends in skill complementarity prior to remote adoption.

We also examine two alternative specifications to ensure robustness to different baseline constructions. The main specification includes three pre-treatment lead terms. The first alternative specification includes only a single lead term, while the second includes two lead terms. Across all three models, the pre-treatment coefficients remain statistically insignificant. This consistency indicates that our findings are not sensitive to the choice of baseline period and provides additional support for the validity of our DID identification strategy.

Table E2. Timing Falsification Test

| | Skill_ _Complementarity (1) |
|---|---|
| Pre(-1) x Treatment | 0.001 |
| | (0.001) |
| Pre(-2) x Treatment | 0.001 |
| | (0.001) |
| Pre(-3) x Treatment | -0.001 |
| | (0.001) |
| Remote_Intensity | 0.003* |
| | (0.001) |
| Quarterly FE | YES |
| Firm FE | YES |
| Controls | YES |
| Observations | 29586 |
| Number of Firms | 2286 |

Note: The term *Pre(-i)* × *Treatment* denotes the interaction between the treatment indicator and the period *i* quarters prior to the start of treatment. Periods earlier than three quarters before treatment are omitted and serve as the baseline category. ***p < 0.01, ** p < 0.05, * p < 0.1.

### E.3. Alternative Estimator for Staggered DID

In staggered DID designs, conventional TWFE estimates may be biased when treatment effects differ across adoption cohorts and evolve over time (Baker et al. 2022). This concern is relevant to our setting because firms adopt remote work at different points of time.

As a robustness check, we therefore implement the Callaway and Sant'Anna (2021) estimator, which estimates group-time average treatment effects by comparing each adoption cohort with appropriate not-yet-treated control firms. We then aggregate these cohort-period estimates to obtain the overall average treatment effects (ATT). We use not-yet-treated firms as the comparison group because these firms are likely to provide a more comparable counterfactual than never-treated firms, while also avoiding comparisons with already-treated firms (e.g., Goodman-Bacon 2021).

We apply this alternative estimator to our baseline model and obtain quantitatively similar results, reported in Table E3 below. Therefore, our main findings are unlikely to be driven by biases associated with staggered treatment timing or heterogeneous treatment effects under the conventional TWFE specification.

Table E3. Alternative Robust Estimation of Impact of Remote Work on Skill Complementarity

| | Skill_Complementarity | |
|---|---|---|
| | (1) | (2) |
| ATT | 0.003* | 0.003* |
| | (0.002) | (0.002) |
| Quarterly FE | YES | YES |
| Firm FE | YES | YES |
| Controls | NO | YES |
| Observations | 13530 | 13530 |
| Number of Firms | 1143 | 1143 |

Notes: This table reports robustness results using the Callaway and Sant'Anna (2021) difference-in-differences estimator for staggered treatment adoption. Firms are grouped into cohorts based on the first quarter in which they adopt remote work. The estimator constructs group-time average treatment effects and aggregates them to obtain the overall Average Treatment Effect on the Treated (ATT). ***$p < 0.01$, ** $p < 0.05$, * $p < 0.1$.

**Appendix References**

Alabdulkareem, A., Frank, M. R., Sun, L., AlShebli, B., Hidalgo, C. A., & Rahwan, I. (2018). Unpacking the polarization of workplace skills. *Science Advances,* 4(7), eaao6030.

Baker, A. C., Larcker, D. F., & Wang, C. C. Y. (2022). How much should we trust staggered difference-in-differences estimates? *Journal of Financial Economics,* 144(2), 370–395.

Callaway, B., & Sant'Anna, P. H. C. (2021). Difference-in-differences with multiple time periods. *Journal of Econometrics,* 225(2), 200–230.

DiMaggio, P. J., & Powell, W. W. (1983). The iron cage revisited: Institutional isomorphism and collective rationality in organizational fields. *American Sociological Review,* 48(2), 147–160.

Goodman-Bacon, A. (2021). Difference-in-differences with variation in treatment timing. *Journal of Econometrics,* 225(2), 254–277.

Leary, M. T., & Roberts, M. R. (2014). Do peer firms affect corporate financial policy? *The Journal of Finance,* 69(1), 139–178.

Neffke, F. M. H. (2019). The value of complementary co-workers. *Science Advances,* 5(12), eaax3370.

Neffke, F., Otto, A., & Weyh, A. (2017). Inter-industry labor flows. *Journal of Economic Behavior & Organization,* 142*, 275–292.*

Shue, K. (2013). Executive networks and firm policies: Evidence from the random assignment of MBA peers. *The Review of Financial Studies, 26(6),* 1401–1442.